\documentclass[12pt]{JHEP3}
\pdfoutput=1

\usepackage{amssymb}
\usepackage{graphicx}
\usepackage{subfigure}
\usepackage{url}

\makeatletter

\title{Revised cosmological parameters after BICEP\,2 and BOSS}

\author{
Santanu Das, Suvodip Mukherjee \& Tarun Souradeep\\
Inter-University Centre for Astronomy and Astrophysics, Post
Bag 4, Ganeshkhind, Pune 411007, India \\
E-mail: 
\email{santanud@iucaa.ernet.in},
\email{suvodip@iucaa.ernet.in},
\email{tarun@iucaa.ernet.in}
}

\abstract{
Estimation of parameters of the \lq standard\rq \,model of cosmology have dramatically improved over past few decades due to 
increasingly exquisite measurements made by Cosmic Microwave Background (CMB) experiments. Recent data from 
Planck matches well with the minimal $\Lambda$CDM model. A likelihood analysis using Planck, WMAP and a selection of high resolution experiments (highL),  
tensor to scalar ratio $r_{0.002}$ is found to be $<0.11$ when $dn_{s}/d\ln k = 0$. Planck also imposes an upper bound on neutrino mass $\sum m_\nu<0.23\,$eV 
using Planck+WMAP+highL+BAO likelihood. However, recently results from 
BICEP 2 claims the detection of $r= 0.2^{+0.07}_{-0.05}$ from polarization spectra. Further, results 
from SDSS-III BOSS large scale galaxy survey constrains the total neutrino mass to $\sum m_\nu=0.36 \pm 0.10$ eV. It is important to study the  consequences of these new measurements on other cosmological parameters. In this paper we  
assess the  revised constraints on cosmological parameters in light of these two measurements that are in some tension with the constraints from Planck. 

The sensitivity of Planck to weak lensing effect on the CMB angular power spectrum suggests that the normalized amplitude of physical lensing power $A_L>1$ at $2\sigma$ hints at a potentially important internal inconsistency.  Therefore, we also include a study of the constraints on $A_L$. Using the prior on $\sum m_\nu$ as measured by SDSS-III BOSS and BICEP 2 likelihood, 
we find that the model with running spectral index ($dn_{s}/d\ln k \neq 0$) leads to a value of $A_L>1$ at $3.1 \sigma$. But, the model with $dn_{s}/d\ln k =0$ makes $A_L$ consistent with 
$1$, at $2.1\sigma$ 
 and also shows that $N_{\rm eff}$ is consistent with its theoretical value of $3.046$ at around $2\sigma$. 
Therefore, the analysis in this paper shows that 
the model with $dn_{s}/d\ln k =0$ gives consistency with  other cosmological parameters ($N_{\rm eff}$ and $A_L$ ) when the current limits on $\sum m_\nu$ and $r_{0.05}$ are considered.  However, on reducing the value of  $r_{0.05}$, the model with non-zero $dn_{s}/d\ln k$ gives consistent result of $A_L =1$.
}
\maketitle

\begin{document}
 
Cosmic Microwave Background (CMB) is a very powerful probe for improving
our understanding of the Universe. Several CMB missions like COBE,
WMAP, Planck, ACT, SPT, BICEP etc., have ushered an era of precision
cosmology. High resolution CMB ESA space mission Planck \cite{Planck_1}
have measured CMB power spectra match extraordinarily 
well with the minimal $\Lambda$CDM model (in particular at angular scales smaller
than few degrees corresponding to multipole $l>50$). Planck has also pinpointed the allowed range of
several cosmological parameters  from the temperature
spectra alone \cite{Planck_param}. Further improvement on these 
constraints are expected in the final release that would include CMB polarization data. However, 
recent results from BICEP 2 \cite{BICEP_1} claimed
the detection of $C_{l}^{BB}$ spectra arising due to the primordial
Gravitational Wave (GW) with tensor to scalar ratio $(r)=0.2_{-0.05}^{+0.07}$,
which is in mild tension with Planck \cite{Planck_param}, which estimates $r<0.11$ without
running spectral index ($dn_{s}/d\ln k=0$). However, this tension reduces while considered 
running spectral index $(dn_{s}/d\ln k\neq0)$ 
in the temperature power spectra \cite{Planck_param}. Also the constraint on neutrino mass $(\sum m_{\nu}=0.36\pm0.10$ eV)
using Baryon Oscillation Spectroscopic Survey (BOSS) CMASS data release
11 by Beutler et al.\cite{SDSS_1} 
is in mild tension with the measurement of Planck ($\sum m_\nu <0.23$ eV). %Therefore, the constraint on neutrino mass ($\sum m_\nu <0.23$ eV) from Planck is in mild tension with the measurement of BOSS. Imposing prior constraint
On considering together the measurement of $r_{0.05}$ and  $\sum m_{\nu}$ can significantly change the current limits on
best-fit $\Lambda$CDM parameters obtained by Planck \cite{Planck_param}.

Since the release  of two prime experiments BICEP 2 and BOSS, several authors, like 
\cite{Archidiacono2014,Zhang2014,Zhang2014a,Cheng2014,Cheng2014a,Cheng2014b,xia,Wu,jerome} have placed constraints on different cosmological parameters using the available data. 
In this paper, we %are interested in 
study the effects of high value 
of $r_{0.05}$ claim by BICEP 2 \cite {BICEP_1} and high non-zero $\sum m_{\nu}$ claim from BOSS-CMAS \cite{SDSS_1} on other cosmological
parameters, using 
%cosmological parameter estimation code 
SCoPE \cite{scope}. Since, tensor contribution to temperature power spectra
is at low multipoles $l <100$, the detection of large tensor contribution
to $C_{l}^{TT}$ is revealed by the amplitude difference between the CMB power at the Sachs-Wolfe
plateau and that at the acoustic peaks. 
%Constraining lower value of $r_{0.05}$ from $C_{l}^{TT}$ implies over estimating the scalar perturbation. 
Therefore, change in $r_{0.05}$ leads to a significant variation of several cosmological parameters which are constrained by the plateau and acoustic peaks. On the other hand 
significant  neutrino mass changes the relativistic matter content of the universe at early times and hence 
%incorporating the constraint on 
%neutrino mass higher than the $0.23eV$, we want to find the allowed range of
it affects the parameters like, $H_{0}$, $\Omega_{b}h^{2}$, 
$\Omega_{c}h^{2}$, $N_{\rm eff}$, $A_{L}$, that depend on the matter 
content of the universe and perturbations in them.
%which are highly degenerate with neutrino mass $\sum m_{\nu}$.
% In this paper 

%We present a comprehensive study of cosmological parameters
%by including the limits on $r_{0.05}$ and $\sum m_{\nu}$ using cosmological parameter estimation code SCoPE \cite{scope}. 
We make an extensive study of the effects
on various parameters and search for a model that remains valid in 
our current theoretical regime. This improved constraint on the cosmological 
parameters are important 
for a better understanding of our cosmological models in light of the  BICEP 2
and BOSS. 

 The paper is organized as follows, in Sec. \ref{cp} we present the estimation
of cosmological parameters  from WMAP-9, Planck and BICEP 2 data using the likelihood
provided by them. Discussions and conclusions of the paper are provided in Sec. \ref{con}.
\begin{figure}
\centering 
\subfigure[]{ \includegraphics[trim=3.2cm 9cm 3.2cm 9.5cm, clip=true, width=0.35\textwidth]{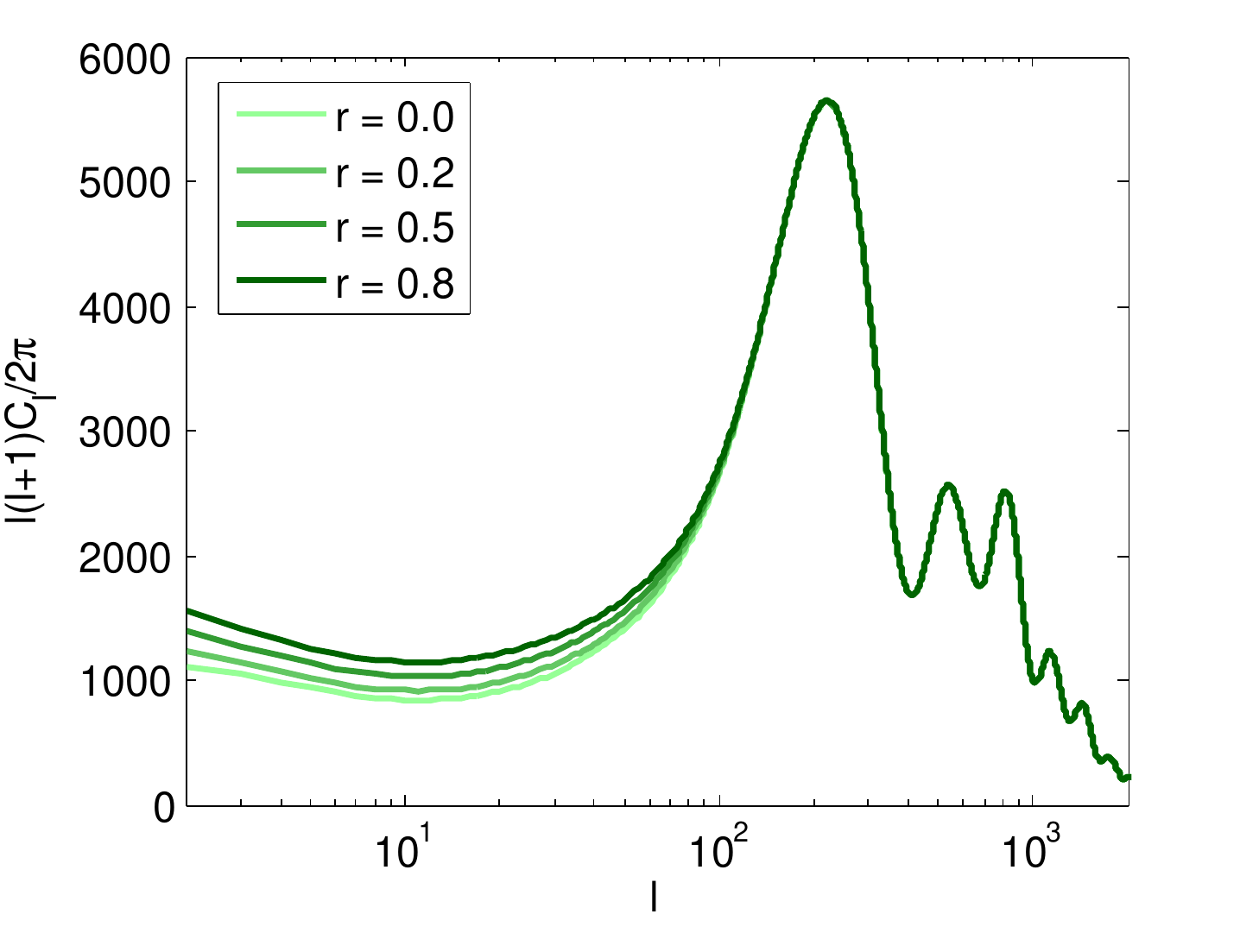} \label{fig:Varyr} }
\subfigure[]{ \includegraphics[trim=3.2cm 9cm 3.2cm 9.5cm, clip=true, width=0.35\textwidth]{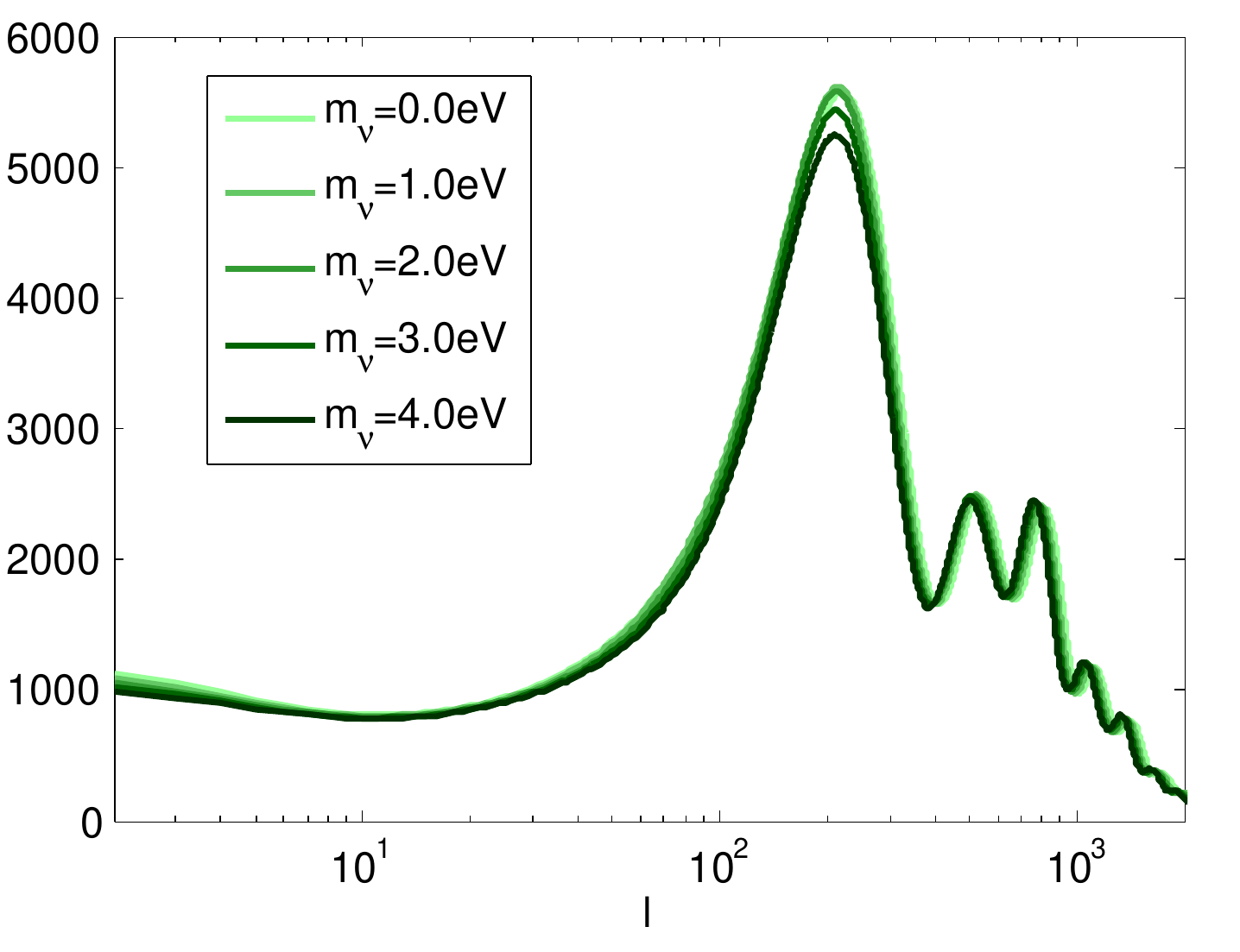} \label{fig:Varymnu}}
\subfigure[]{ \includegraphics[trim=3.2cm 9cm 3.2cm 9.5cm, clip=true, width=0.35\textwidth]{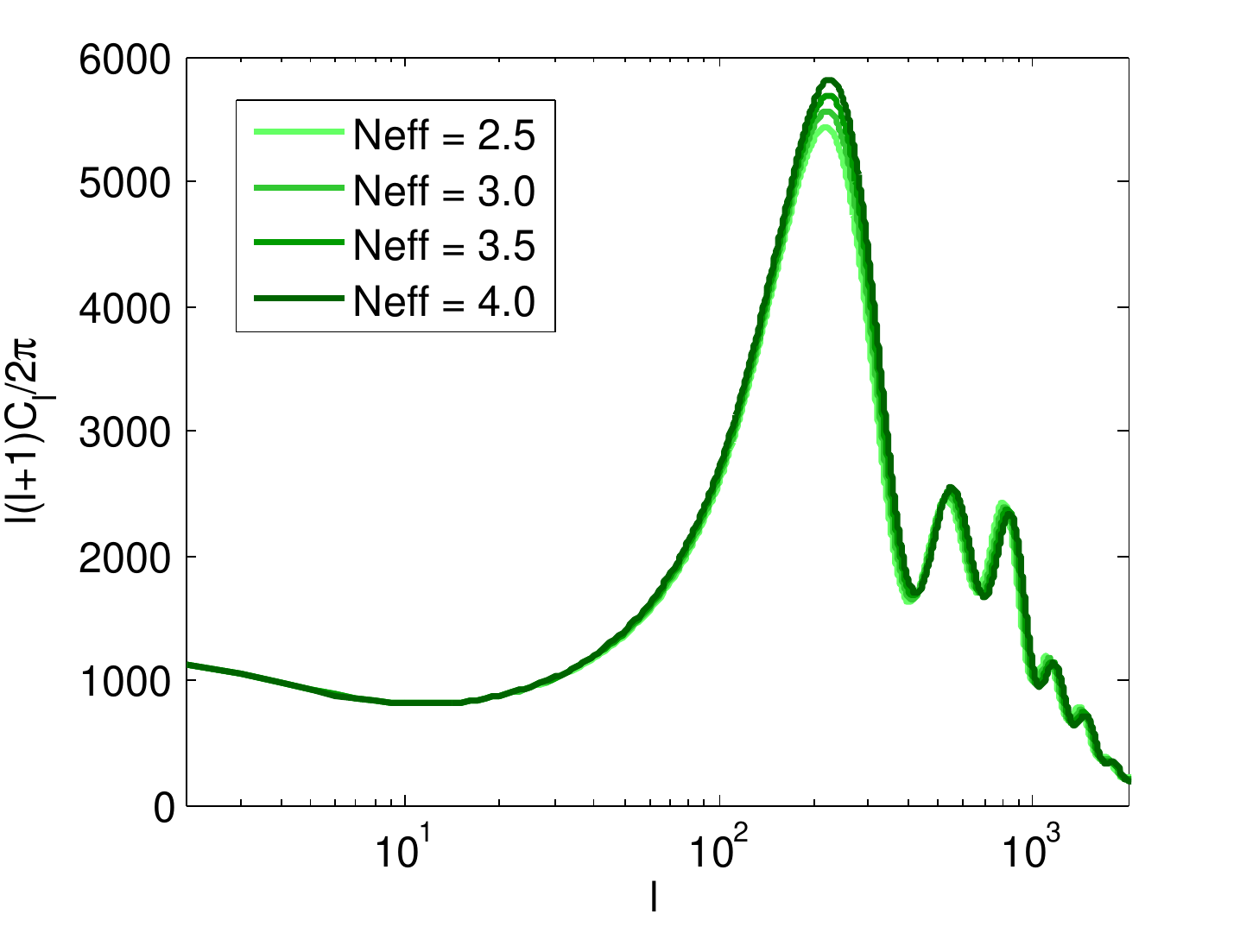} \label{fig:VaryNeff}}
\subfigure[]{ \includegraphics[trim=3.2cm 9cm 3.2cm 9.5cm, clip=true, width=0.35\textwidth]{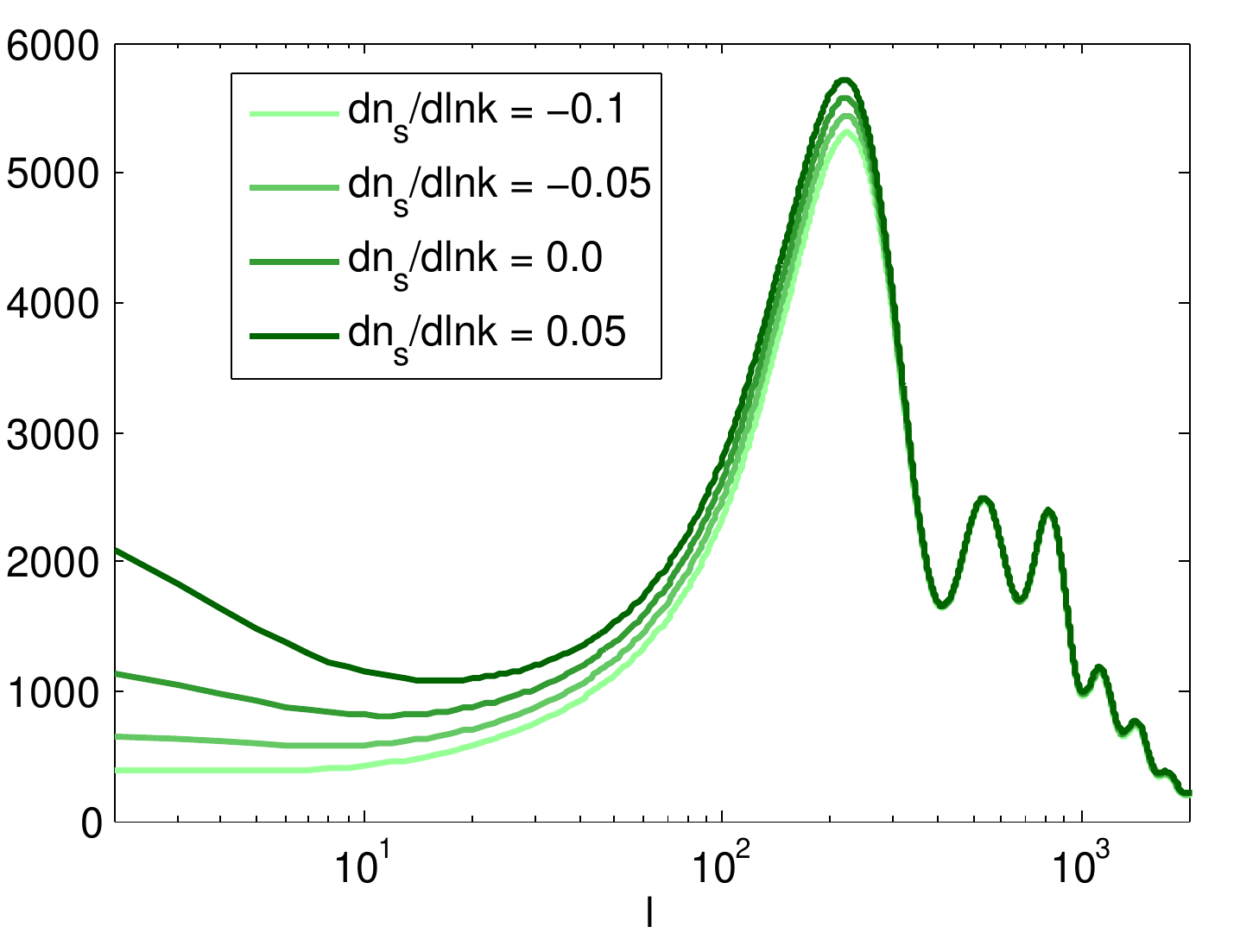} \label{fig:Varydnsdk}}
\caption{Temperature power spectra, $l(l+1)C_{l}^{TT}/2\pi$ for different sets of $r, \sum m_\nu, N_{eff}$ and $dn_s/d\ln k$ values obtained using CAMB \cite{camb}. Other parameters are kept fixed at their best-fit $\Lambda$CDM values \cite{Planck_param}.}\label{fig:1} 
\end{figure}
% \begin{figure}[H]
% \centering
% \includegraphics[width=5in,keepaspectratio=true]{Neff.eps}
% \caption{\label{fig:nef}Temperature power spectra, $l(l+1)C_{l}^{TT}/2\pi$ for different sets of $dn_s/d\lnk$ values with other best-fit $\Lambda$CDM model generated using CAMB\cite{camb}.}
% \end{figure}
% \caption{Temperature power spectra, $\frac{l(l+1)C_{l}^{TT}}{2\pi}$ for different sets of $dn_s/dlnk$ values with other best-fit $\Lambda$CDM model generated using CAMB\cite{camb}.}\label{fig:ns_cl} 

%In Fig. \ref{fig:1}, we plot the effect of different values of $r_{0.05}$ and $\sum m_\nu$ on temperate 
%power spectra. affects 

\section{Cosmological Parameter Estimation}\label{cp}

%In this section 
We calculate the constraints on different cosmological parameters using
WMAP-9, Planck and BICEP 2 likelihood. However, before going to the main analysis, 
it is important to understand the effect of different parameters like $r, 
\sum m_\nu, N_{eff}$ and $ dn_s/d\ln k$ on CMB angular power spectrum. 
%temperature power spectra $C_l^{TT}$, we plot their effects 
Fig.~\ref{fig:1} shows the effects of different parameters on $C_l^{TT}$. 
Fig.~\ref{fig:Varyr} shows the effect of %different values 
$r_{0.05}$ on $C_l^{TT}$ for a range of values. All the other cosmological parameters
are kept fixed at their standard values and we use $n_t=0$. As the tensor 
power spectrum is dominant only at low multipoles, 
only lower multipoles get affected due to this variation. Plots show that $r_{0.05}$ 
is expected to be correlated with $n_s$, as that %$n_s$ 
also affects the low multipole power. Fig.~\ref{fig:Varymnu}
and Fig.~\ref{fig:VaryNeff} 
show the effect of neutrinos on $C^{TT}_l$. 
Change in either of $N_{\rm eff}$ or $\sum m_{\nu}$,
affects the total matter fraction in the universe, which leads to change in the 
expansion history of the universe and shifts the  
CMB peaks towards larger or smaller scales. 
These also affect the epoch of matter radiation equality and hence the ratio between 
the even and the odd peaks in the CMB power spectrum changes. 
In Fig.~\ref{fig:Varymnu} we show the effects of variation of $\sum m_{\nu}$ 
keeping $N_{\rm eff}$ fixed at $3.048$, whereas in Fig.~\ref{fig:VaryNeff} we choose the neutrinos to be massless. 
Finally, in Fig.~\ref{fig:Varydnsdk} we show the effects of running spectral index 
$dn_s/d\ln k$ on $C_l^{TT}$. 
%whose effect found to be constrained in the 
%lower multipoles (~ upto 300). 

%In this paper, we discuss
We present most of the results using the following likelihoods \-:- 
1) For Planck+WP + BICEP 2 likelihood, 
where we add up the results from commander\_v4.1\_lm49.clik,
lowlike\_v222.clik, CAMspec\_v6.2TN\_2013\_02\_26.clik likelihood
\cite{wmap_like,Planck_like} and BICEP 2 likelihood \cite{bicep_like} to perform parameter estimation. 
2) For Planck+WP+ BICEP 2 + Lensing likelihood, where we add the lensing 
likelihood along with the other likelihoods. %The average value of the nuisance parameters as mentioned by Planck 
%\cite{Planck_param} are used in this paper for all the cases.

We vary the Standard Six Parameters (SP),
$\{\Omega_{b}h^{2}$, $\Omega_{m}h^{2}$, $h$, $\tau, n_{s}$, $A_{s}\}$, along with
the other parameters like $dn_s/d\ln k$, $n_t$, $N_{\rm eff}$, $A_L$. The 
ranges of priors used for all these parameters are provided in Table. \ref{tab_1}.
%As we are interested in finding the implications on the cosmological parameters due to the recent results from BICEP 2 and SDSS-III BOSS, we use the probability density function (pdf) %one dimensional likelihood for $r_{0.05}$ and $\sum m_{\nu}$, from these experiments as the prior in our MCMC analysis while estimating the parameters.  
We estimate parameters for the cosmological models, 
(a) SP, (b) SP + either of $\{dn_{s}/d\ln k$, $n_{t}$, $N_{eff}\}$ and (c) SP + 
$\{A_L +dn_s/d\ln k \}$ for all different sets of $\sum m_\nu$ and $r_{0.05}$ mentioned below.
\begin{enumerate}
\item
$\sum m_\nu=0$ and $r=0$, (WP + Planck likelihood and WP + Planck + Lensing likelihood).
\item
$\sum m_\nu \neq 0$ and $r=0$, (WP + Planck likelihood and WP + Planck + Lensing likelihood).
\item
$\sum m_\nu=0$ and $r_{0.05} \neq 0$, (WP + Planck +BICEP 2 likelihood and WP + Planck + Lensing +BICEP 2 likelihood).
\item
$\sum m_\nu \neq 0$ and $r_{0.05} \neq 0$, (WP + Planck + BICEP 2 likelihood and WP + Planck + Lensing + BICEP 2 likelihood).
\end{enumerate}

\begin{table}[h]
\centering
\caption{Prior range for the Cosmological parameters used in this paper.}
\label{tab_1} 
\begin{tabular}{|l|l|p{8cm}|}
%\begin{tabular}{|clc|cl|p{8cm}|}
\hline 
Parameters  & Prior range  & Defination \tabularnewline
\hline 
\hline 
$\Omega_{b}h^{2}$  & 0.015 $\textendash$  0.03 & Physical baryon density. \tabularnewline
\hline 
$\Omega_{m}h^{2}$  & 0.05 $\textendash$ 0.25 & Physical matter density. \tabularnewline
\hline 
$h$ & 0.55 $\textendash$ 1.2 & Hubble parameter. \tabularnewline
\hline 
$\tau$ & 0.01 $\textendash$ 0.2 & Reionization optical depth. \tabularnewline
\hline 
$n_{s}$ & 0.8 $\textendash$ 1.2 & Scalar spectral index. \tabularnewline
\hline 
$\ln (10^{10}A_{s})-2\tau$ & 2.5 $\textendash$ 3.2 & Amplitude of temperature fluctuations. \tabularnewline
\hline 
$r_{0.05}$  & 0 $\textendash$ 1 & Ratio of tensor primordial power to curvature power at $k_{0}=0.05\, \rm Mpc^{-1}$. \tabularnewline
\hline 
$dn_{s}/d\ln k$  & -2.0 $\textendash$ 2.0 & Running of the scalar spectral index. \tabularnewline
\hline 
$n_{t}$ & -0.25 $\textendash$ 0.25 & Tensor spectral index. \tabularnewline
\hline 
$N_{\rm eff}$  & 1.0 $\textendash$ 5.0 & Effective number of nutrino-like relativistic degrees of freedom. \tabularnewline
\hline 
$\sum m_{\nu}$  & pdf from BOSS \cite{SDSS_1}& The sum of neutrino masses . \tabularnewline
\hline 
$A_{L}$  & 0.5 $\textendash$ 3.5 & Amplitude of the lensing power relative to the physical value. \tabularnewline
\hline 
% \captionof{table}{Cosmological parameters used in this paper.}\label{tab_1}
%\tabularnewline
% \captionof{table}[t]{Cosmological parameters used in this
%paper.}\label{tab_1} 
\end{tabular}
\par
\end{table}
\begin{center}
%\label{tab_1} 
\par\end{center}

%\subsection{Six parameters with $r=0.2$ and $\sum m_{\nu}$.}
%The tensor contribution to temperature power spectra, $D_{l}^{TT}=l(l+1)C_{l}^{TT}/2\pi$
%dominate only at the lower value of $l<100$ and then decreases rapidly
%with $l$. Due to this, the dominant effect of tensor perturbation
%on $D_{l}^{TT}$is in the low $l$. As a result, any change in the
%tensor contribution to temperature spectra, changes the height difference
%between the $1^{st}$ peak and the plateau region as shown in the
%Fig. \ref{fig:1}. This slight increase in power at low $l$ for $r=0.2$
%can significantly change several parameters like $\tau$, $dn_{s}/d\text{ln k}$,$n_{s}$,
%$h$ etc. From the measurement of $r_{0.002}=0.2$ from BICEP 2 \cite{BICEP_1}
%which is higher (consistent) with the measurement $r_{0.002}<0.11$
%$(r_{0.002}<0.26)$ from Planck \cite{Planck_param} for $dn_{s}/d\text{ln k}=0$
%$(dn_{s}/d\text{ln k}\neq0)$. We are interested to find the improved
%constraints on different cosmological parameters. For the SP model,
%we want to see the allowed regions of these $6$ parameters with $dn_{s}/d\text{ln k}=0$.
%This model is an alternative to the running spectral index parameters
%mentioned by Planck \cite{Planck_param}. 

\subsection{Beyond Standard $6$ parameters with $\sum m_\nu$ and $r_{0.05}$}
\begin{figure}[h]
\centering
\subfigure[]{
\includegraphics[trim=0.1cm 10.1cm 0.3cm 10.2cm, clip=true, width=0.80\columnwidth]{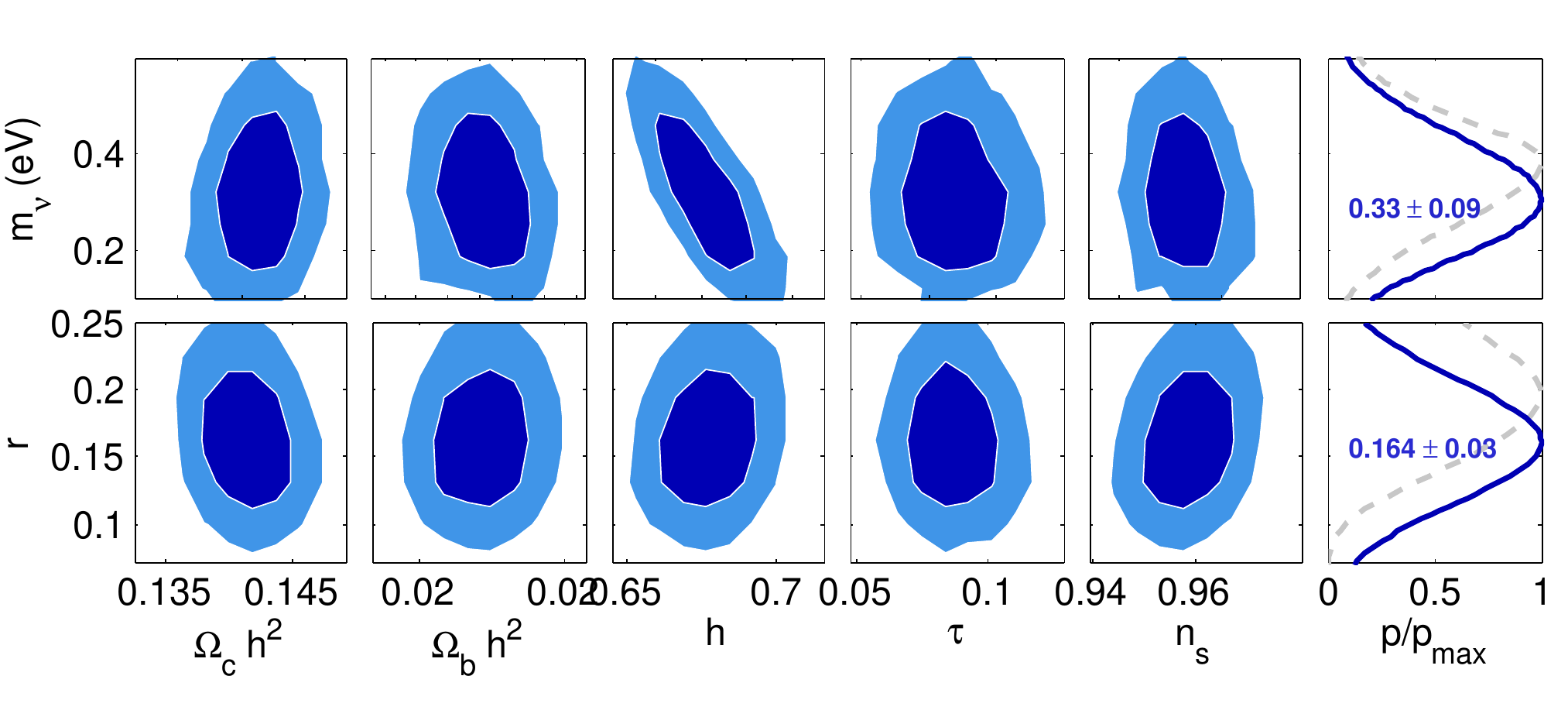}\label{nolen}
}
\subfigure[]{
\includegraphics[trim=0.1cm 10cm 0.3cm 10.2cm, clip=true, width=0.80\columnwidth]{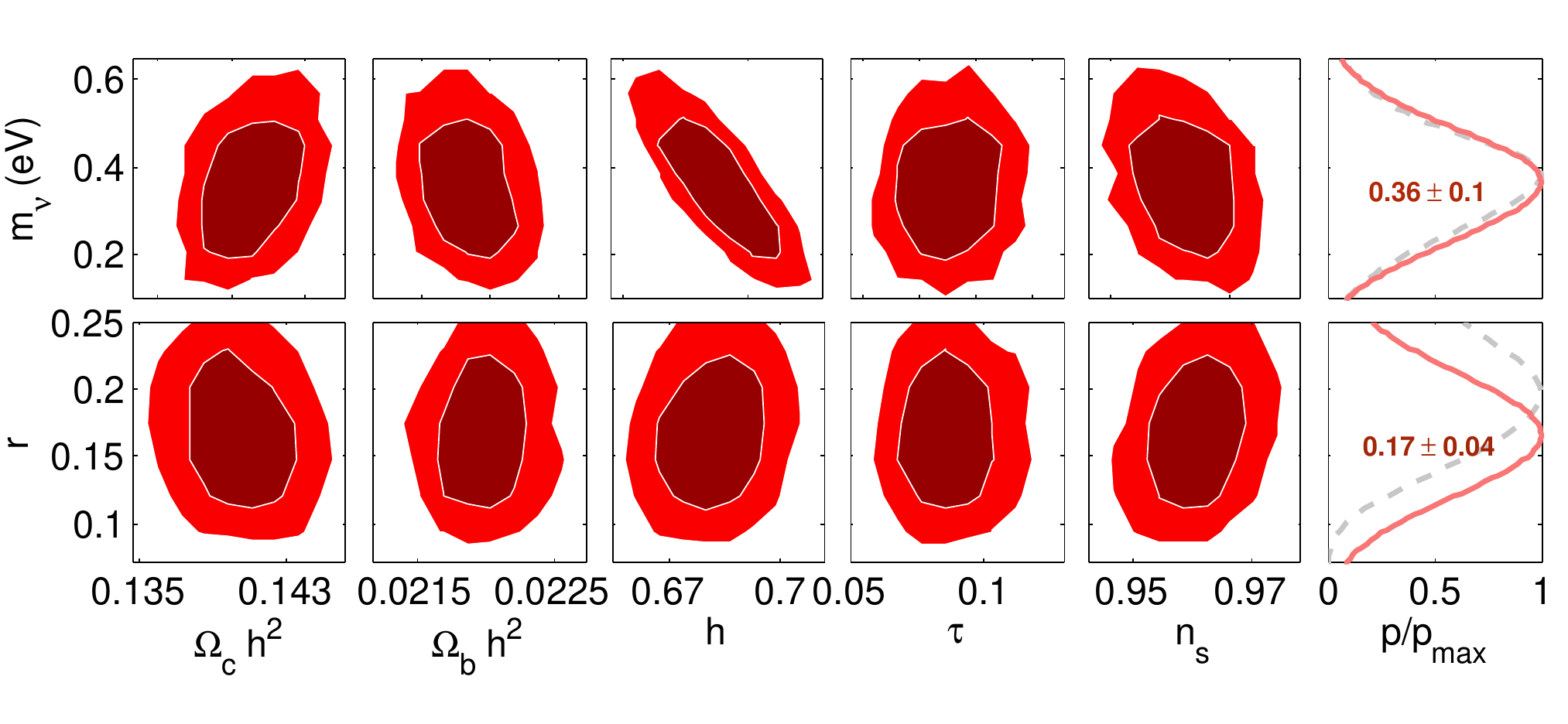}\label{len}
}
\caption{\label{fig:SP_2D}The two dimensional likelihood contours for the
SP with $r_{0.002}$ and $\sum m_{\nu}$. (a): WP + Planck +BICEP 2
likelihood. (b): WP + Planck + lensing +BICEP 2 likelihood. The one dimensional marginal
probability distribution for $r_{0.05}$ and $\sum m_{\nu}$ are shown in
the last column. Gray dotted lines show the posterior on $r_{0.05}$ and $\sum m_{\nu}$ as obtained by
BOSS and BICEP 2 respectively. 
%Planck likelihood is driving $r_{0.05}$ to very low values. 
%However as the prior is very small at small $r_{0.05}$,
%the result can't come down to a very low value. 
%In both the figures we have shown the prior on $r_{0.05}$ only in the range 0.05 to 0.25  and on
%$\sum m_{\nu}$ in the range 0.1 to 0.65 eV. 
}\label{fig:4}
\end{figure}

\begin{figure}[h]
\centering
\includegraphics[trim=0.6cm 7.5cm 0.1cm 7.8cm, clip=true, width=0.78\columnwidth]{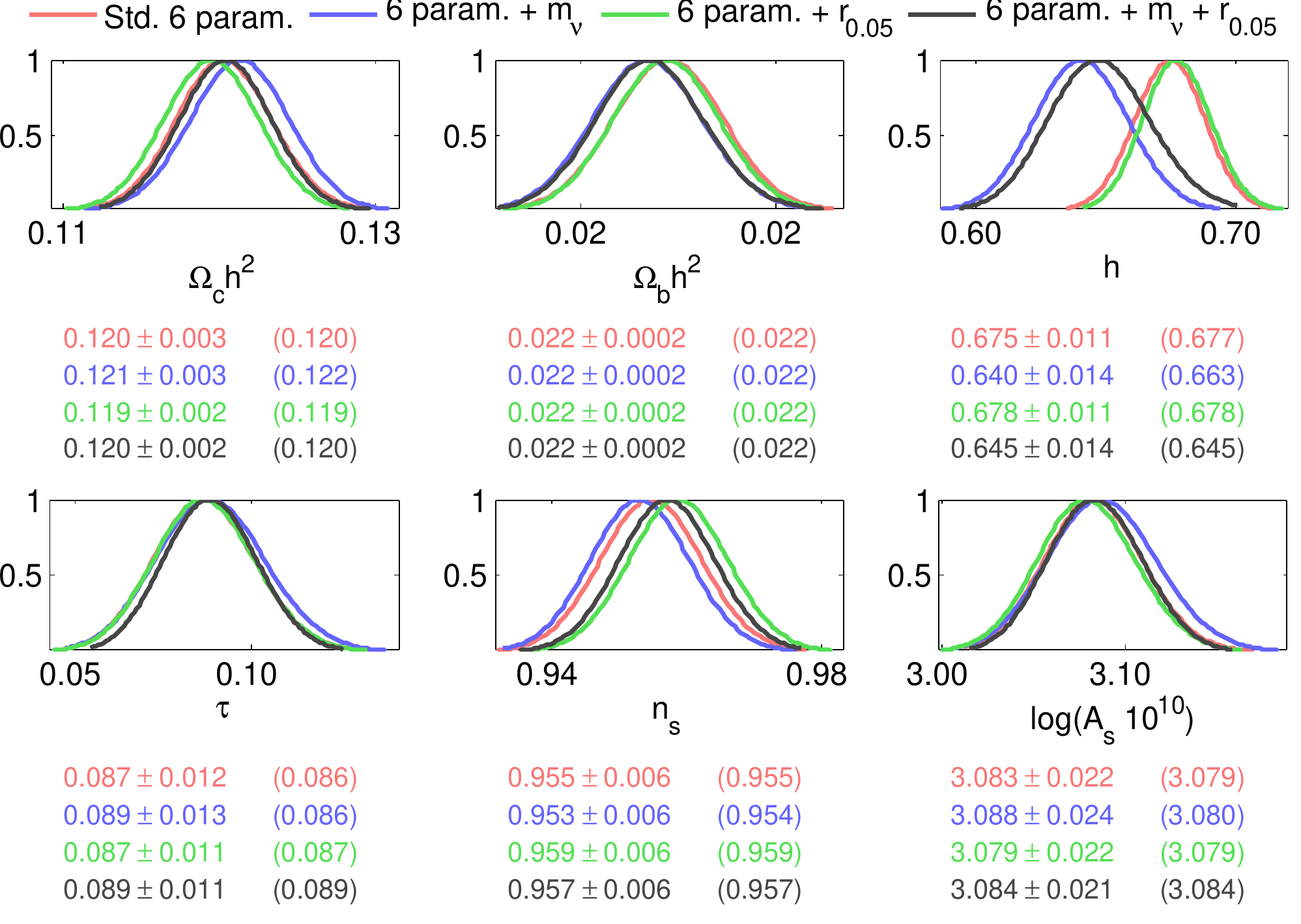} 
\caption{\label{fig:SP_nolensing}One dimensional marginal probability distribution 
of the standard model parameters for four different cases without considering lensing likelihood. Red : Only the standard
model parameters (SP) are varied. The neutrinos are considered to
be massless. Blue : We vary $\sum m_{\nu}$ along with other
SP. The number of massive neutrino species here are considered to
be $N_{\rm eff}=3.046$. For both these models we considered only the
scalar power spectrum. Green : Analysis with $r_{0.05}$ and SP.
Neutrinos are considered to be massless and $n_{t}=0$. Black
: We vary both $\sum m_{\nu}$ and $r_{0.05}$. The average and the
standard deviations for the parameters are given in the plot itself.
The best fit values are quoted in the brackets. }
\end{figure}

\begin{figure}[h]
\centering
\includegraphics[trim=1.3cm 7.9cm 0.1cm 7.7cm, clip=true, width=0.80\columnwidth]{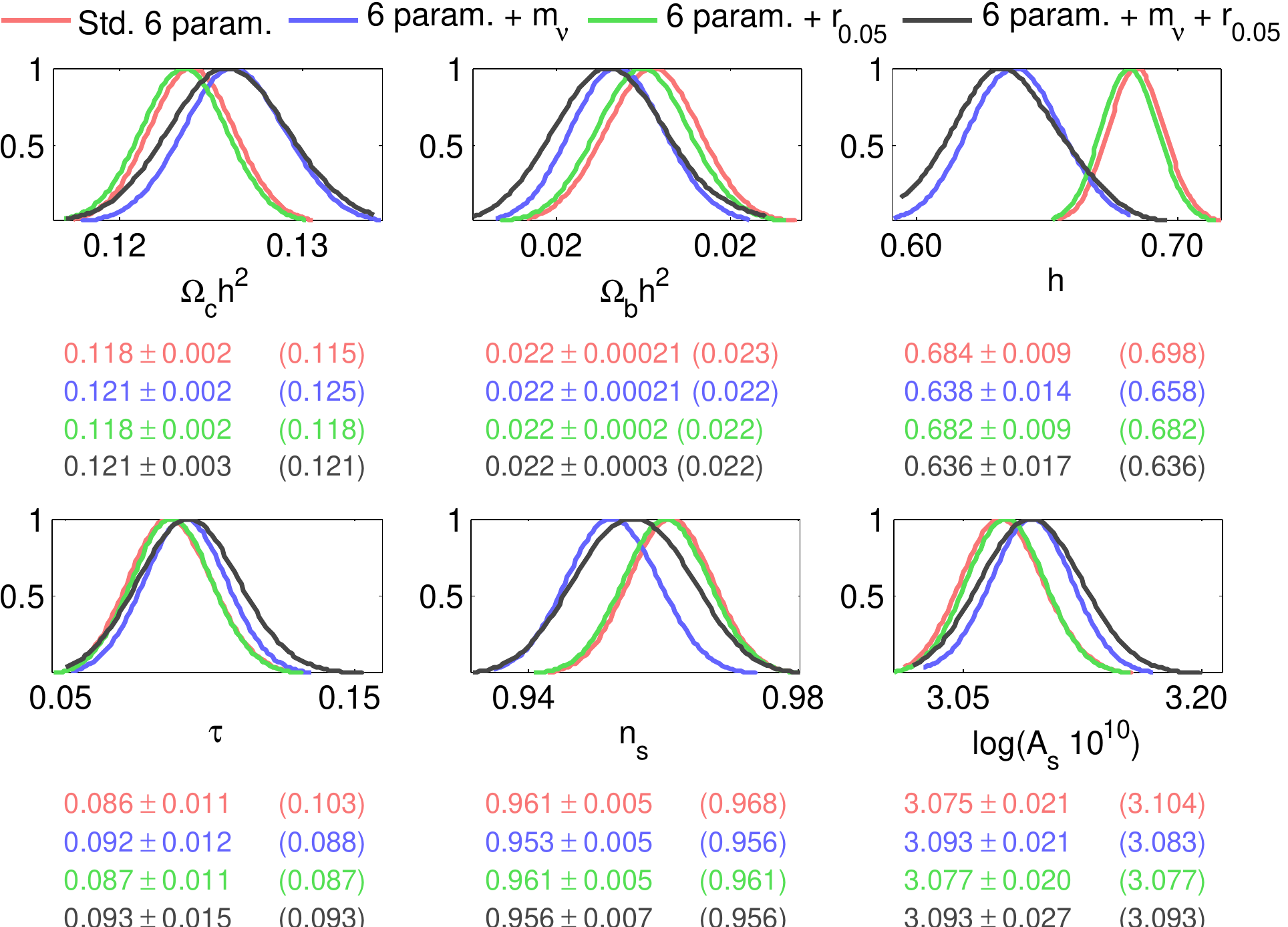} 
\caption{\label{fig:SP_lensing}One dimensional marginal probability distribution 
of the standard model parameters for four different cases after considering lensing likelihood. The results are shown for four different cases. Red :
Only the standard model parameters (SP) are varied. The neutrinos
are considered to be massless here. Blue : We vary $\sum m_{\nu}$
along with other SP. The number of massive neutrino species here
are considered to be $N_{\rm eff}=3.046$. For both these models we considered
only the scalar power spectrum. Green : Analysis with $r_{0.05}$
and SP. Neutrinos are considered to be massless and $n_{t}=0$.
Black : We vary both $\sum m_{\nu}$ and $r_{0.05}$. The average
and the standard deviations for the parameters are given in the plot
itself. The best fit values are quoted in the brackets. }
\end{figure}

First we discuss the statistics of %change in the posterior of the 
standard $6$ parameters $\{\Omega_{b}h^{2}$, $\Omega_{c}h^{2}$, $h$, 
$\tau$, $n_{s}$, $A_{s}\}$
due to the change in $r_{0.05}$ and $\sum m_{\nu}$. 
%As shown in Fig. \ref{fig:1}, higher value of $r_{0.002}$ increases the power at low $l$. This have a significant effect on the best-fit cosmological parameters.
From the plots %The change in $C_{l}$ due to the variation in $m_{\nu}$ is 
shown in Fig. \ref{fig:1}, it is evident that if we
change the value of $r_{0.05}$ or $\sum m_{\nu}$ then there can be significant
effects on the cosmological parameters. We run SCoPE \cite{scope} 
with standard $6$ cosmological parameters for all the cases mentioned in Sec. \ref{cp}, and the results are shown in Fig. \ref{fig:4}, Fig. \ref{fig:SP_nolensing} and Fig. \ref{fig:SP_lensing}. 
%At this point it should be noted that during our MCMC analysis instead of using flat prior on
%$r_{0.05}$ and $\sum m_{\nu}$, we have used the one dimensional likelihood,
%from BICEP 2 and BOSS experiment respectively, as the prior for $r_{0.05}$ and $\sum m_{\nu}$. %As shown in Fig. \ref{fig:4}, %\ref{fig:SP_nolensing},  $\sum m_{\nu}$ have a strong effect on $h$. 
The two dimensional contour plots of $r_{0.05}$ and $\sum m_{\nu}$ with the standard model parameters 
with WP + Planck + BICEP 2 likelihood and WP + Planck + lensing + BICEP 2 likelihood are shown in blue and red color respectively in Fig. \ref{fig:SP_2D}.   It can
be seen that $\sum m_{\nu}$ is strongly correlated with $h$ with
the correlation coefficient $\sim -0.74$. As $\sum m_{\nu}$ increases the
value of $h$ decreases, which is expected as the matter content
of the universe is getting changed and leads to change in the epoch of matter-radiation 
equality of the universe. $\sum m_{\nu}$ also has small effect on the 
baryon density and the dark matter content of the universe. The effect of $r_{0.05}$ on $h$ shows slight positive correlation as shown in  
Fig. \ref{fig:4}. %On including both $r_{0.05}$ and $\sum m_\nu$, the value of $h$ decreases as shown in Fig. \ref{fig:SP_nolensing}, \ref{fig:SP_lensing}.
%than the  on $h$ are opposite, and hence if we %the case which include both $\sum m_\nu$ and $r_{0.05}$, then the change in %leads to much lower value of $h$ is almost null, that can be  seen %as shown in Fig. \ref{fig:SP_nolensing}, \ref{fig:SP_lensing}.
We can also notice that on using only WP + Planck + BICEP 2 likelihood, the average value
of $\sum m_{\nu}$ is $0.33$ eV whereas if lensing is included the average
value increases and becomes $0.36$ eV. Though, as variance is
high, both the values are within $1\sigma$ of each other. The value
of $r_{0.05}$ is same in both the cases and is approximately $0.16$.

%Other parameters are nearly consistent for all the $4$ cases. 

%However, due to the change in $h$, the value of the best-fit parameters
%$\{\Omega_{m},\Omega_{b},\Omega_{\lambda}\}$ significantly changes as mentioned in Eq. \eqref{eq2}. 
%\begin{align}
Fig. \ref{fig:SP_nolensing} and Fig. \ref{fig:SP_lensing} show
the one dimensional probability distribution for the standard cosmological
parameters from WP + Planck+ BICEP 2 likelihood and WP + Planck+ lensing + BICEP 2 likelihood respectively. Including both $m_{\nu}$ and $r_{0.05}$ in the estimation leads to lower value of Hubble constant $(h= 0.645)$ as shown in Fig. \ref{fig:SP_nolensing} and Fig. \ref{fig:SP_lensing}, which is in tension with the results obtained from calibrated SNe magnitude-redshift relation by Riess et al. \cite{snia}. This tension increases further on taking into account of the lensing 
likelihood as shown in Fig. \ref{fig:SP_lensing}.
%\begin{figure}[h]
%\centering
%\includegraphics[trim=0.8cm 7cm 0.6cm 6.5cm, clip=true, width=0.79\columnwidth]{standard6paramnolensed.pdf} 
%nof the standard model parameters when analyzed with Planck+WP likelihood.
%The results are shown for four different cases. Red : Only the standard
%model parameters (SP) are varied. The neutrinos are considered to
%be massless. Blue : we vary $\sum m_{\nu}$ along with other
%SP. The number of massive neutrino species here are considered to
%be $N_{\rm eff}=3.046$. For both these models we considered only the
%scalar power spectrum. Green : analysis with $r_{0.002}$ and SP.
%Neutrinos are considered to be massless and $n_{t}=0$. Black
%: we vary both $\sum m_{\nu}$ and $r_{0.002}$. The average and the
%standard deviations for the parameters are given in the plot itself.
%The best fit values are quoted in the brackets. }
%\end{figure}

%In Fig. \ref{fig:SP_lensing} we show the result by including the 
%lensing likelihoods in our analysis. Though,
%the change in the results are small. The two dimensional likelihoods
%with Planck+WP+lensing %for these cases 
%are shown in the Fig. \ref{fig:SP_2D} in red color. 
%We can see that in case of only WP + Planck case, the average value
%of $\sum m_{\nu}$ is $0.349$ eV whereas if lensing is included the average
%value increases and becomes $0.42$ eV. Though, as variance is
%high, both the values are within $1\sigma$ of each other. The value
%of $r_{0.05}$ is same in both the cases and is approximately $0.16$. 

%Fig. \ref{fig:SP_nolensing}, \ref{fig:4} also shows that $r_{0.05}$ is slightly positively
%correlated with $n_{s}$. $r_{0.05}$ also has slight positive correlation
%on $h$, i.e. increase in $r_{0.05}$ increases the values of $h$ slightly. 

%The result from fig(\ref{fig:SP_nolensing}) diagram also shows that
%when we are allowing to vary $r_{0.05}$, it decreases the value of $\Omega_{c}h^{2}$,
%but a nearly same value for $\Omega_{b}h^{2}$ component in the Universe.
%This is mainly due to the a small contribution of tensor spectra at
%acoustic peak. Due to increase in $r_{0.05}$, there is net increase in power
%at low $l$, which is compensated by the decrease in the power of
%Integrated Sachs Wolfe (ISW) term. This leads to an increase in the
%Hubble parameter ($h$) at low redshift. From the 1-D plot in Fig.
%\ref{fig:2}, shows the Hubble parameter is changed to 
%\begin{align}
%H_{0}=69\pm1.2\,\,\text{km}\, s^{-1}\,\text{Mpc \ensuremath{^{-1}}},\label{eq1}
%\end{align}
% which is consistent within $1\sigma$ with SNe Ia+Cepheids measurement
%by Riess et al. \cite{snia}. This consistency gets worse on taking
%higher mass of $\sum m_{\nu}$ as discussed in Sec. \ref{neutrino}.

%The reason for this is mention that section. WRITE MORE.\\
 %Results obtained in this section indicates the changes in standard$\Lambda$CDM model due to just change in tensor to scalar ratio $r_{0.05}$ and $dn_{s}/d\text{ln k}=0$. In the next sections we discuss the effect of $r_{0.002}=0.2$ on other cosmological parameters.
%\subsection{Effect of $r_{0.05}$ and $\sum m_\nu$ on $dn_{s}/d\ln k$ and $n_t$.}\label{ns}
\subsection{Standard $6$ parameters + $dn_{s}/d\ln k$ with $\sum m_\nu$ and $r_{0.05}$}\label{ns}
%In this section, we discuss the effect on running spectral index $dn_{s}/d\ln k$ and $n_t$ 
%due to $r_{0.05}$. In Fig. \ref{fig:Varydnsdk} we plot the effect of change in $dn_{s}/d\ln k$ on $C_l^{TT}$. 

Simplest inflationary models predict that the running of the spectral index $dn_{s}/d\ln k$
is related to the higher order of inflationary slow roll parameters
\cite{kosowsky}. Tight observational constraints on $dn_{s}/d\ln k$, can rule out several inflationary models. Planck \cite{Planck_param}
has put constraint on $dn_{s}/d\ln k$ with $r<0.26$ as, $dn_{s}/d\ln k=-0.022\pm0.010$, $(68\%;\mathrm{Planck+WP+highL})$

%\begin{equation}
%dn_{s}/d\ln k=-0.022\pm0.010  \,\,\,\,\,\, (68\%;\mathrm{Planck+WP+highL})
%\label{eq3}
%\end{equation}

\begin{figure}[h]
\includegraphics[trim=3.4cm 9.0cm 0.3cm 9cm, clip=true, width=0.3\columnwidth]{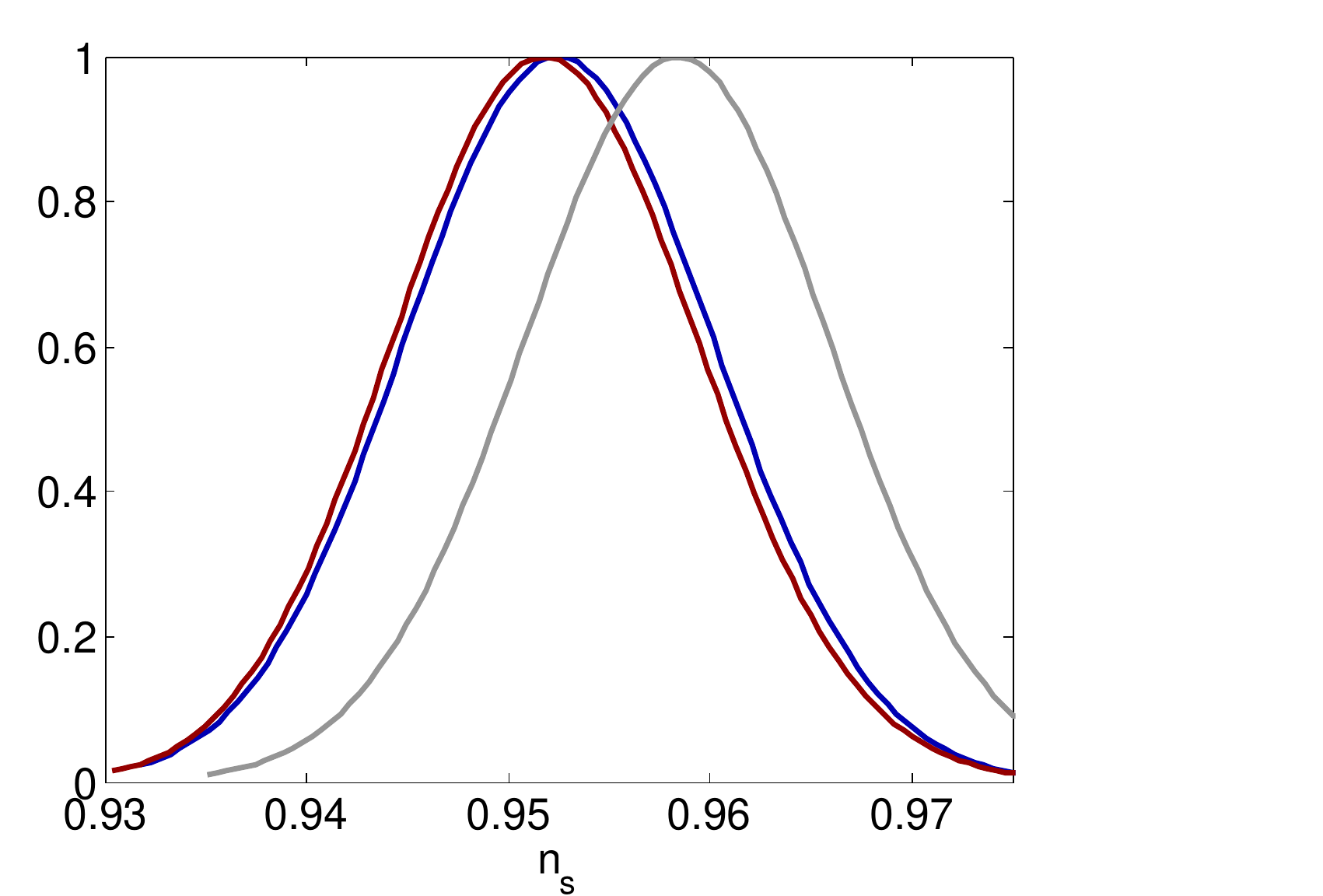}
\includegraphics[trim=5cm 0cm 0.3cm 21cm, clip=true, width=0.50\columnwidth]{top.pdf}
\includegraphics[trim=3.5cm 9cm 1.4cm 9.1cm, clip=true, width=0.28\columnwidth]{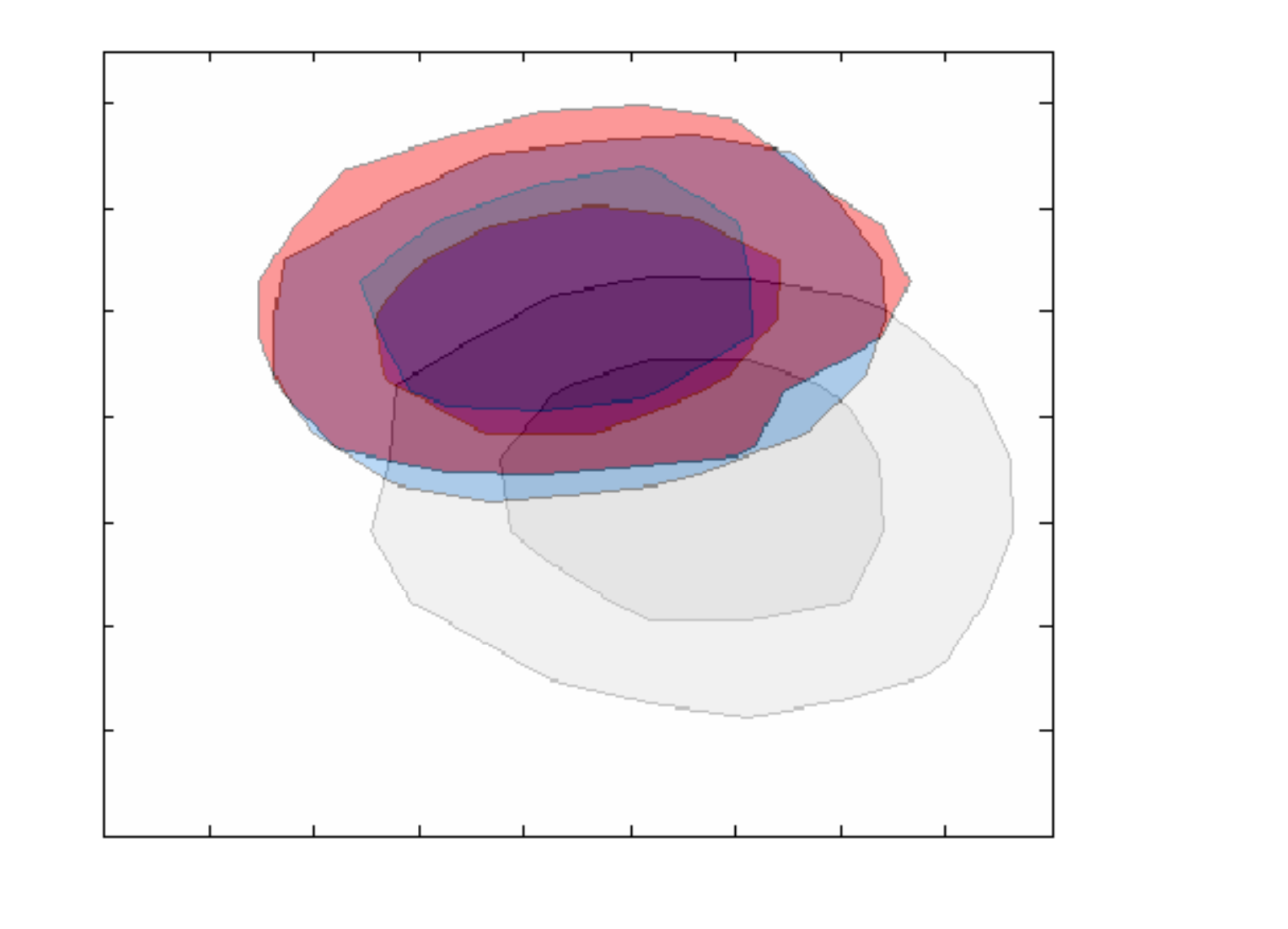}
\includegraphics[trim=3cm 9cm 3cm 8cm, clip=true, width=0.26\columnwidth]{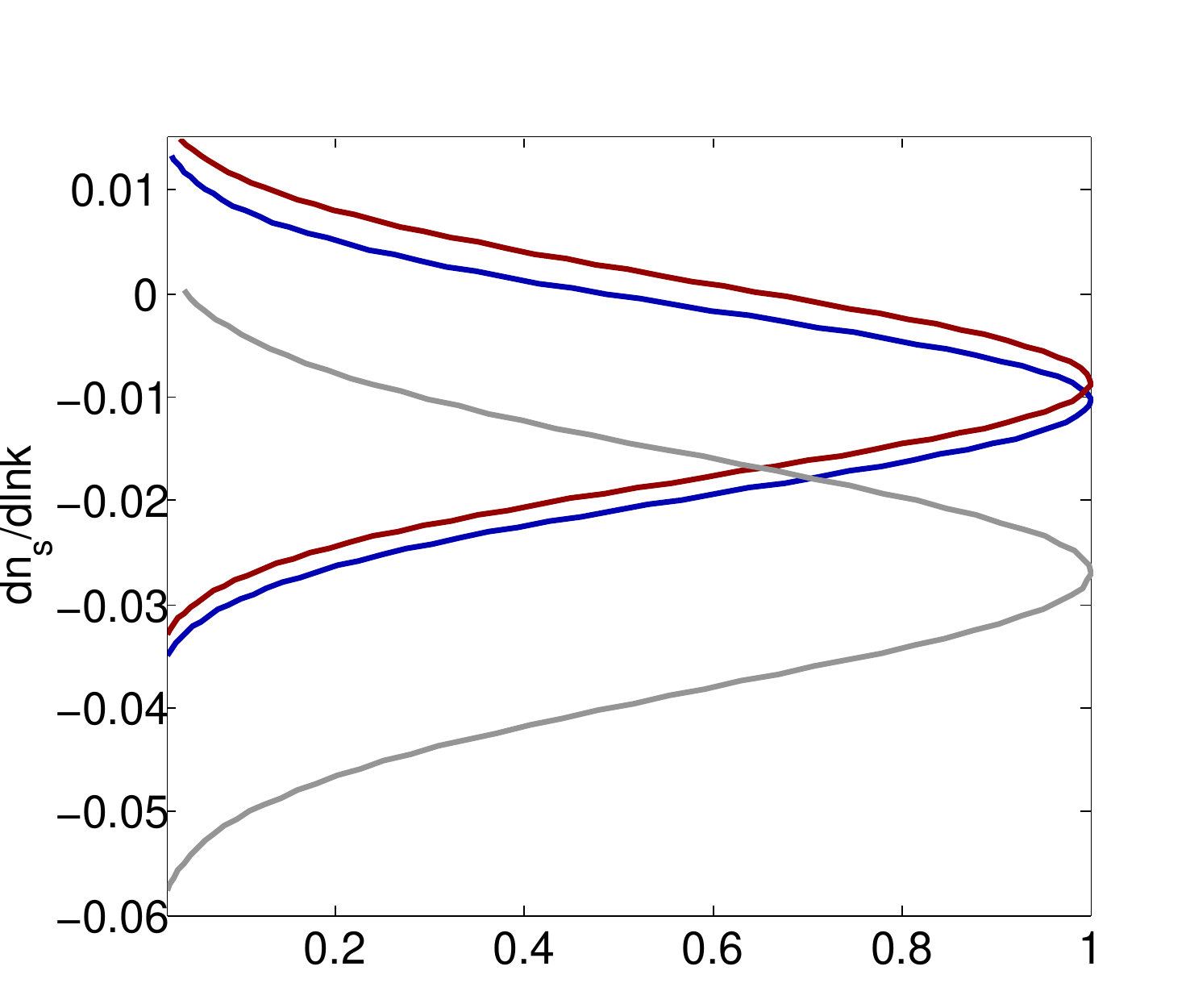} 
%}
%\subfigure[]{
%\includegraphics[width=3.5in,keepaspectratio=true]{dnsdk1D.eps} 
%}
\caption{\label{fig:dnsdk} Variation of $dn_{s}/d\ln k$ and $n_s$ are shown for following cases. Blue : For SP+$dn_{s}/d\ln k$, Red : For
SP+$dn_{s}/d\ln k$+$\sum m_{\nu}$, Gray : for SP+$dn_{s}/d\ln k$+$r{}_{0.05}$.
Introducing tensor part leads to $dn_{s}/d\ln k$
away from the $0$ at $\sim 3.23 \sigma$ .}
\end{figure}

We obtain the constraints on $dn_{s}/d\ln k$ and $n_{s}$ in Fig. \ref{fig:dnsdk}. On considering no tensor spectrum, i.e. $r=0$, 
$dn_{s}/d\ln k$ is well consistent with zero, which also matches Planck results.  However, the constraint on $dn_{s}/d\ln k$ with BICEP 2 likelihood \cite{BICEP_1} shows $3.23\sigma$ deviation from zero. 
The %This shows that the 
effect of $\sum m_\nu$ over  $dn_{s}/d\ln k$ is negligible. The case with both $\sum m_\nu$ and $r_{0.05}$, shows that $dn_{s}/d\ln k$ is consistent with zero at $2.7 \sigma$. We  summarize the results for $n_s$ and $dn_{s}/d\ln k$ on considering the measurement from BICEP 2 and BOSS 
 in Table \ref{tab_2}.

%\begin{figure}[H]
%\centering 
%\includegraphics[width=4in,keepaspectratio=true]{dnsdk.eps}
%\includegraphics[width=6in,keepaspectratio=true]{mnu.eps}
%\end{figure}
\begin{table}[h]
\centering
\caption{Constriants on $n_s$ and $dn_{s}/d\ln k$ for different set of parameters.}\label{tab_2}
\begin{tabular}{|c|c|c|c|c|}
\hline 
 & SP+$\frac{dn_{s}}{d\ln k}$ & SP+$\frac{dn_{s}}{d\ln k}$+$\sum m_{\nu}$ & SP+$\frac{dn_{s}}{d\ln k}$+$r_{0.05}$ & SP+$\frac{dn_{s}}{d\ln k}$+$\sum m_{\nu}$+$r_{0.05}$\tabularnewline
\hline 
\hline 
$n_{s}$ & 0.953$\pm$0.006 & 0.0952$\pm$0.006 & 0.958$\pm$0.006 & 0.9570$\pm$0.0063\tabularnewline
\hline 
$\frac{dn_{s}}{d\ln k}$ & -0.0105$\pm$0.0069 & -0.0083$\pm$0.0072 & -0.0271$\pm$ 0.0084 & -0.0246$\pm$0.0091\tabularnewline
\hline 
\end{tabular}
\end{table}

%\subsubsection{Standard $6$ parameters + $n_{t}$.}\label{nt}
%\begin{figure}
%\centering
%\includegraphics[trim=1.0cm 7.5cm 1.5cm 7.5cm, clip=true, width=0.85\columnwidth]{rnsnt.pdf}
%\caption{\label{fig:ns_r_nt}Constraints on $r_{0.05}$, $n_{s}$ and $n_{t}$. We use the probability distribution of $r_{0.05}$ obtained from BICEP \cite {BICEP_1} as the prior on $r_{0.05}$ for making this estimation. The constraint on $n_{t}$ is poor from Planck+WP data. The tensor pivot scale is chosen at $0.002\,{\rm Mpc}^{-1}$.}\label{fig:nt}
%\end{figure}

\subsection{Standard $6$ parameters + $N_{\rm eff}$ with $\sum m_\nu$ and $r_{0.05}$}\label{neff}
\begin{figure}[h]
\centering
\includegraphics[trim=0.3cm 8.5cm 0.0cm 8.0cm, clip=true, width=0.7\columnwidth]{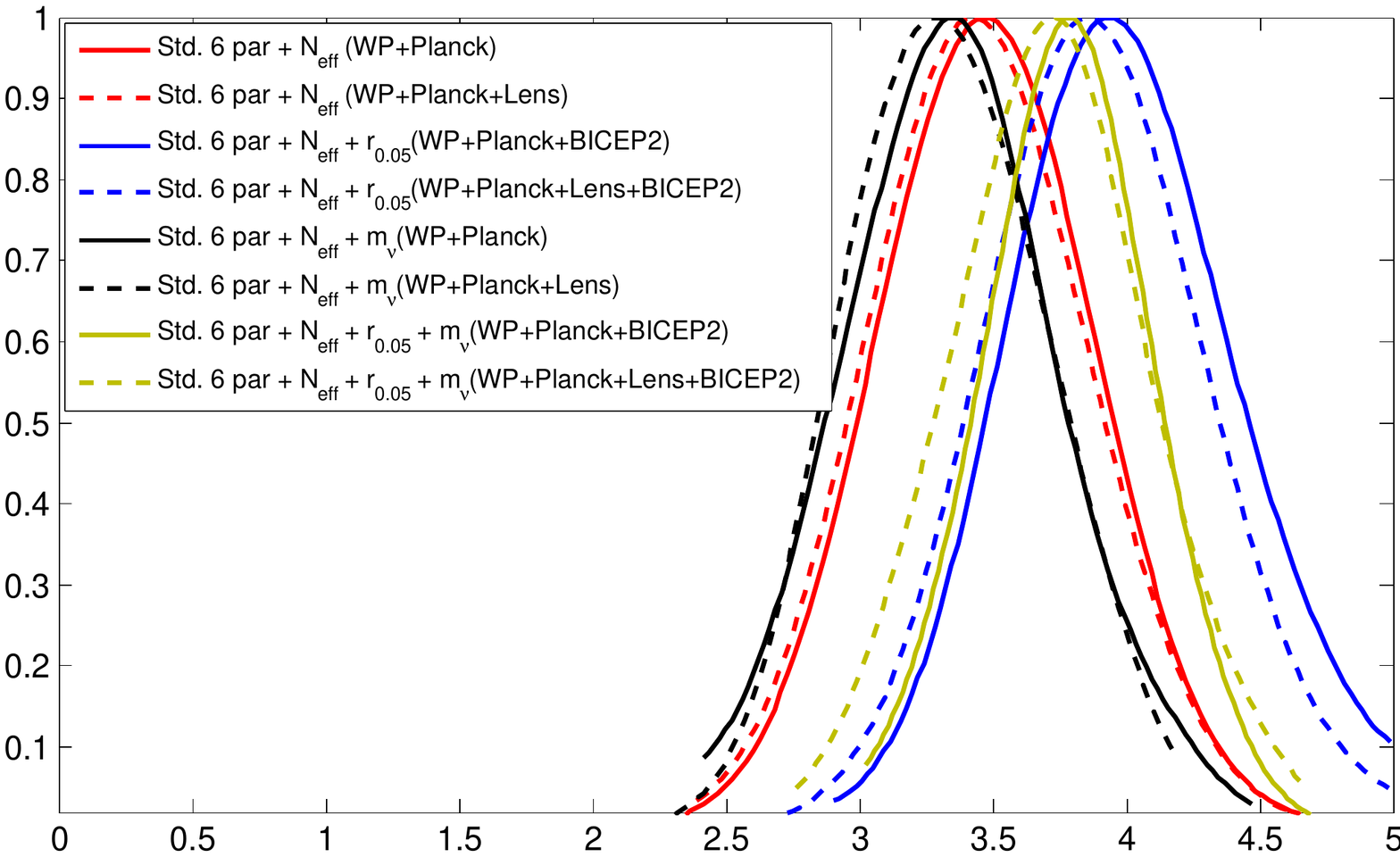}
\caption{\label{fig:Neff_1D}Distribution of $N_{\rm eff}$ for different sets of parameters.}%Planck+WP and Planck+WP+lensing likelihood.}
\end{figure}

%In this section, we put constraints on number of neutrino species in the 
%universe $N_{\rm eff}$ due to the recent limits on $\sum m_\nu$ and $r_{0.05}$. 
In the standard model of particle physics there are  3 types of 
neutrinos corresponding to the three families of leptons. However, there are corrections to the $N_{\rm eff}$ due to non instantaneous decoupling 
and QED effects. This theoretically leads to $N_{\rm eff}= 3.046$. However, any other non-interacting relativistic species also effect the CMB power spectra in the same manner as neutrinos. Signature of any such relativistic species can therefore 
be found by estimating $N_{\rm eff}$ from the CMB measurements.
% because of its effect on CMB power spectrum 
%as shown in Fig. \ref{fig:VaryNeff}. Estimating $N_{\rm eff}$ from 
%Planck data can reveal the presence of 
%any other relativistic species which can behave as neutrinos. 
In Fig. \ref{fig:Neff_1D} we plot the one dimensional likelihood
of $N_{\rm eff}$ for four different cases, (a) $N_{\rm eff}$ with standard 6 parameter case with 
$\sum m_\nu =0$ and $r=0$. (b) $N_{\rm eff}$ with standard 6 parameter, $r_{0.05}$ and 
$\sum m_\nu =0$. (c) $N_{\rm eff}$ with standard 6 parameter case with $\sum m_\nu \neq 0$ and $r=0$. 
(d) Non-zero value for both $\sum m_\nu$ and $r_{0.05}$. All these cases are studied by 
considering both lensing and without lensing.

As shown in the Fig. \ref{fig:Neff_1D}, if the massive neutrinos
are considered then the $N_{\rm eff}$ decreases, whereas if tensor modes
are considered with the massless neutrinos then the $N_{\rm eff}$ of
the neutrinos increases.  The values of $N_{\rm eff}$ 
for different cases are shown below,

\[ N_{\rm eff} = 
\left\{
\begin{array}{l
 l}
3.3221_{\,}\pm0.35 & \quad SP+N_{\rm eff}\,+\sum m_{\nu}\,\rm (Planck+WP)\\
3.2938_{\,}\pm0.32  & \quad SP+N_{\rm eff}\,+\sum m_{\nu}\,\rm (Planck+WP+lensing)\\
3.4867_{\,}\pm0.31  & \quad SP+N_{\rm eff}\,\rm  (Planck+WP)\\
3.4447_{\,}\pm0.34 & \quad SP+N_{\rm eff}\,\rm (Planck+WP+lensing)\\
3.9837_{\,}\pm0.36  & \quad SP+N_{\rm eff}+r_{0.05}\,\rm (Planck+WP+BICEP \,2)\\
3.8867_{\,}\pm0.34  & \quad SP+N_{\rm eff}+r_{0.05}\,\rm (Planck+WP+lensing+BICEP \,2)\\
3.7530_{\,}\pm0.26  & \quad SP+N_{\rm eff}+r_{0.05}+\sum m_{\nu}\,\rm (Planck+WP+BICEP \,2)\\
3.6832_{\,}\pm0.30  & \quad SP+N_{\rm eff}+r_{0.05}+\sum m_{\nu}\,\rm (Planck+WP+lensing+BICEP\, 2)
\end{array}
\right.\]
%}

% \begin{equation}
% N_{eff}=\begin{cases}
% \begin{array}{c}
% \end{array} & \begin{array}{c}
% \end{array}\end{cases}
% \end{equation}
%If we vary both $r_{0.05}$and $\sum m_{\nu}$ then we get $\bf N_{eff}=3.2923\pm0.3349$ (with lensing) and $\bf N_{eff}= 3.41\pm0.32$ (without lensing)

\begin{figure}[h]
\centering
\includegraphics[trim=0.0cm 9.0cm 0.0cm 9.0cm, clip=true, width=0.85\columnwidth]{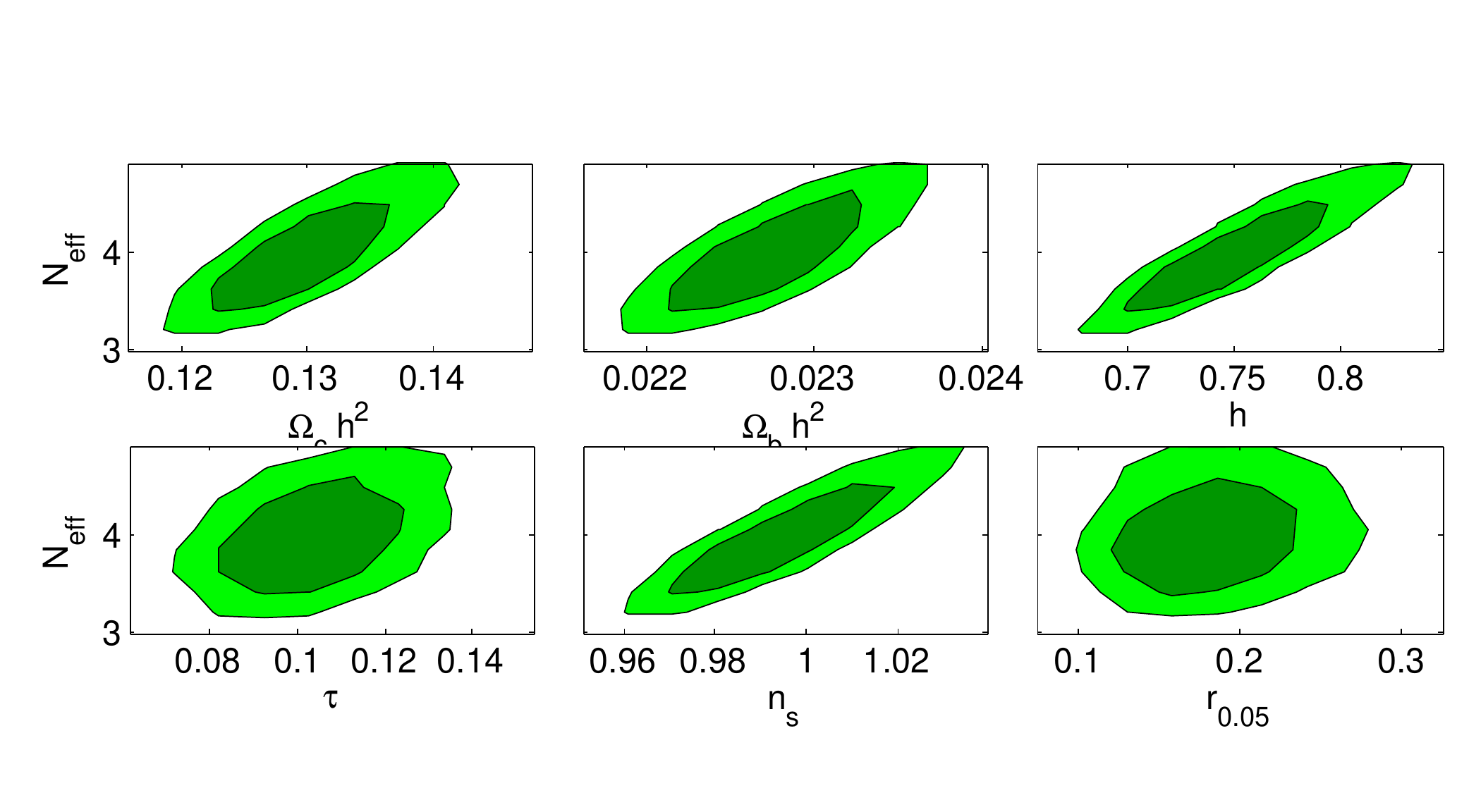}
\caption{\label{fig:Neff_r}Two dimensional likelihood contours of $N_{\rm eff}$ for Planck+WP+BICEP 2 likelihood 
with standard model parameters and $r_{0.05}$. Neutrinos are considered
to be massless for this case.}
\end{figure}

\begin{figure}[h]
\centering
\includegraphics[trim=0.0cm 9.0cm 0.0cm 9.0cm, clip=true, width=0.75\columnwidth]{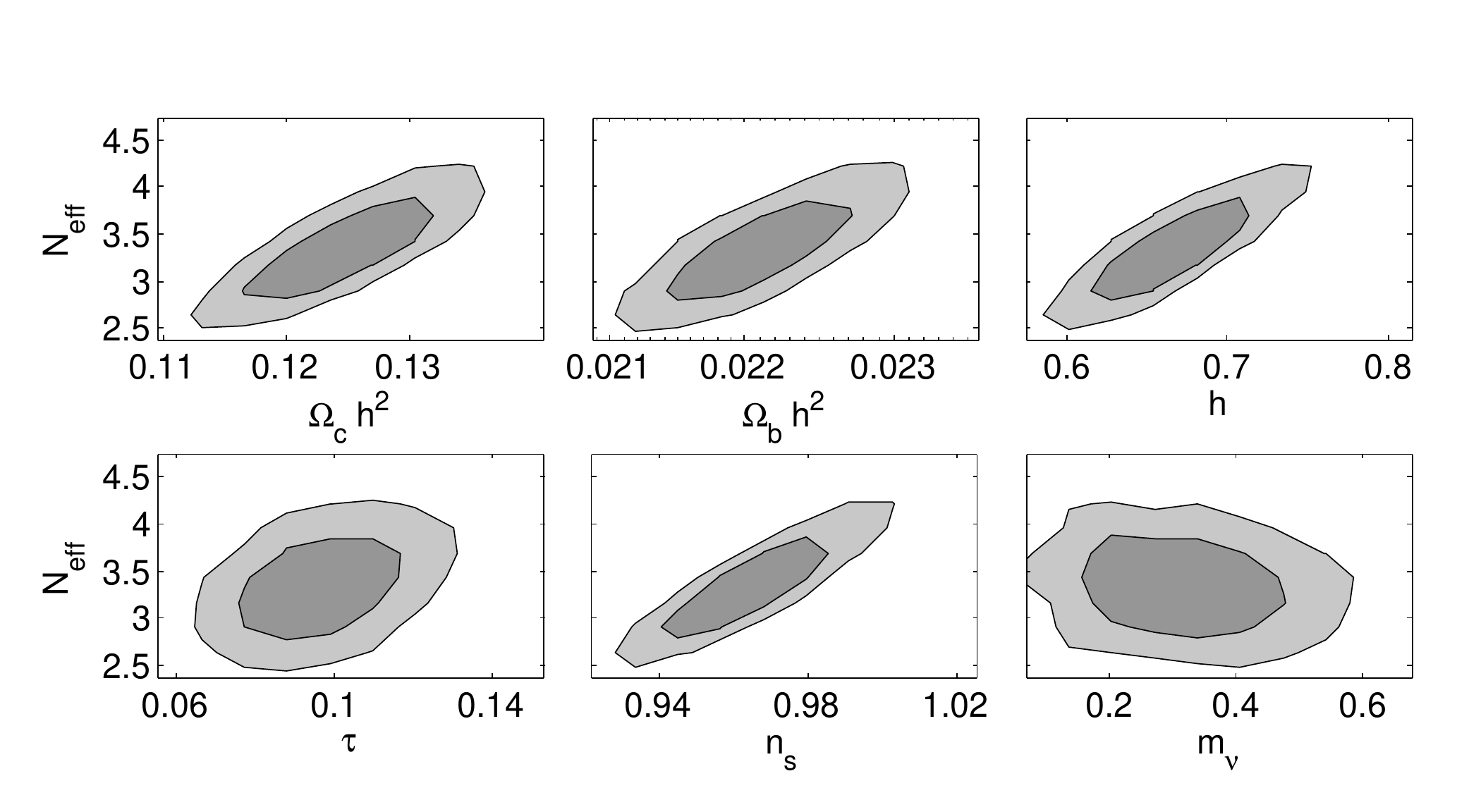}
\caption{\label{fig:Neff_mnu}Two dimensional likelihood contours of $N_{\rm eff}$ for Planck+WP likelihood
with standard model parameters and $\sum m_{\nu}$. We have considered $r=0$ for this case. }
\end{figure}
In Fig. \ref{fig:Neff_r}, we plot the two dimensional likelihood
of $N_{\rm eff}$ with all the other parameters for massless neutrino
case, which shows that $N_{\rm eff}$ is positively correlated
with almost all other parameters. % In 
Fig. \ref{fig:Neff_mnu} shows the two dimensional likelihood with $\sum m_\nu \neq 0$ and $r=0$. 
%For these figures (Fig. \ref{fig:Neff_r} and Fig. \ref{fig:Neff_mnu}), we do not consider lensing likelihood. 
\subsection{Standard $6$ parameters + $A_{L}$ + $dn_{s}/d\ln k$ with $\sum m_\nu$ and $r_{0.05}$}\label{al}
%with $\sum m_{\nu}=0.36\, eV$ and $r=0.2$.}

\begin{figure}
\centering
\subfigure[]{
\includegraphics[trim=0.0cm 8.0cm 0.0cm 7.5cm, clip=true, width=0.45\columnwidth]{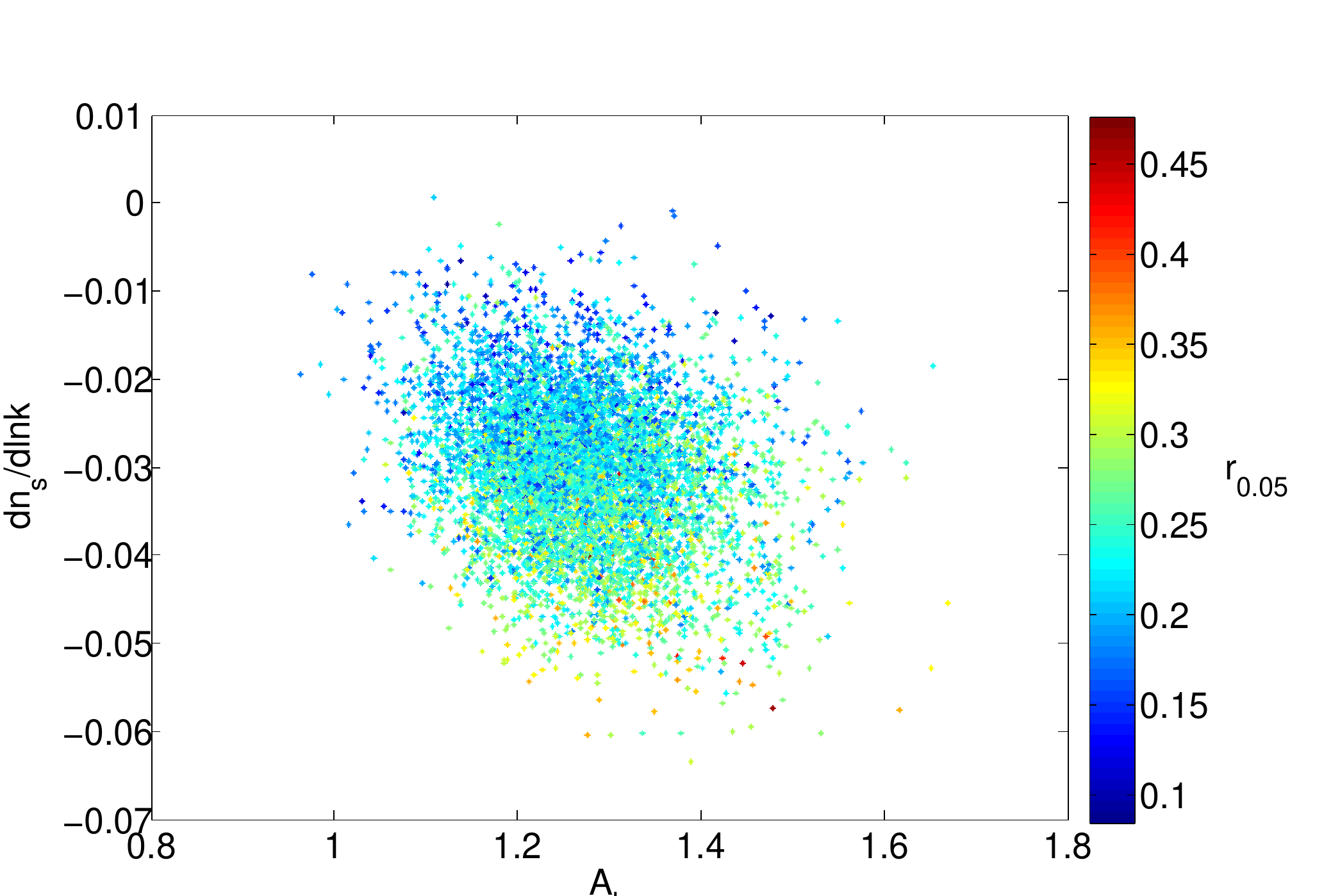}
}
\subfigure[]{
\includegraphics[trim=0.0cm 8.0cm 0.0cm 7.5cm, clip=true, width=0.45\columnwidth]{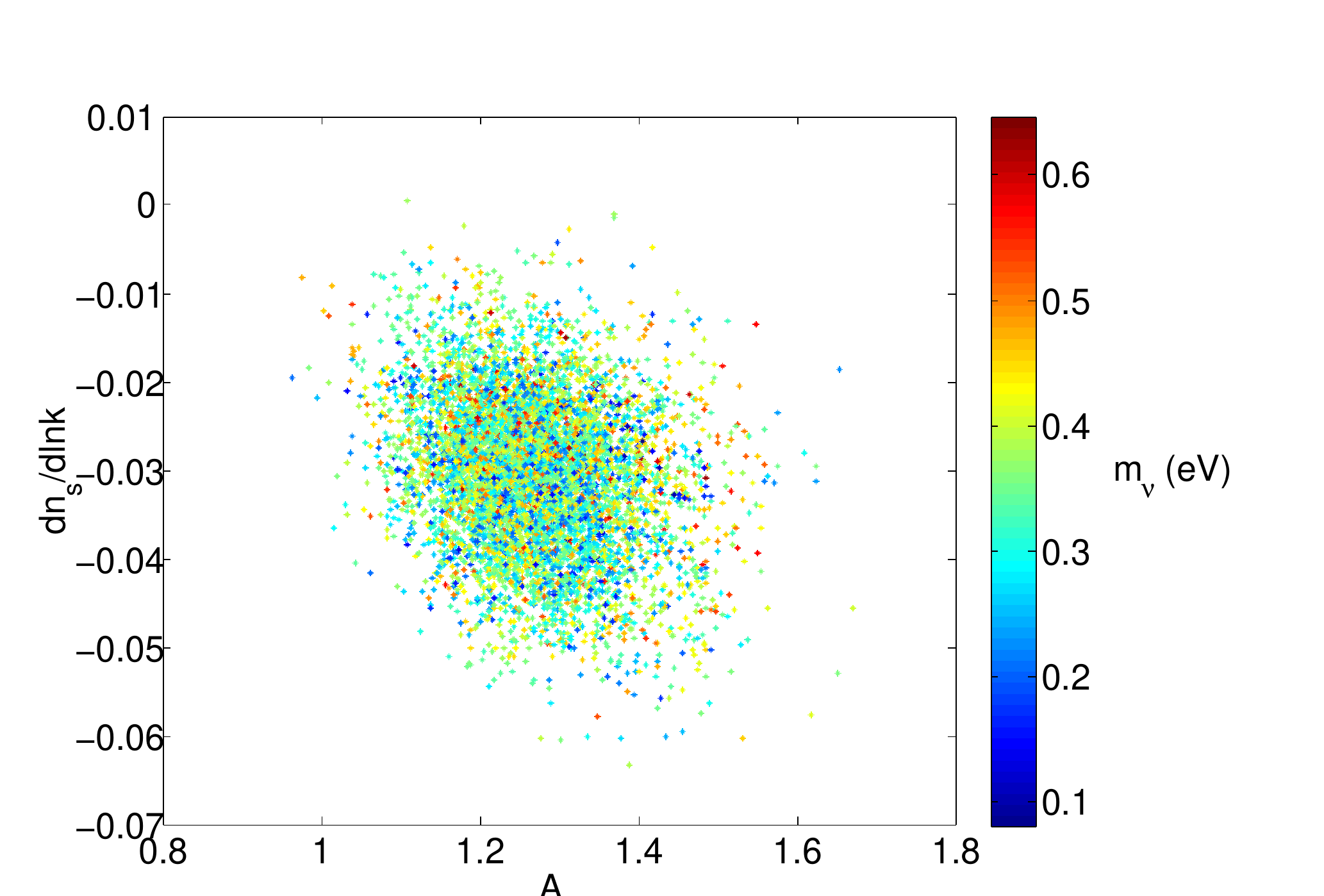}
}
\caption{\label{fig:scatteredplot} We plot $A_{L}$ vs $dn_{s}/d\ln k$
in the scattered diagram and colour coded it with (a) $r_{0.05}$ and (b) $\sum m_{\nu}$.
$A_{L}$ is positively correlated with both $r_{0.05}$ and $\sum m_{\nu}$
and is negatively correlated with $dn_{s}/d\ln k$. }
\end{figure}

\begin{figure}[h]
\centering
\includegraphics[trim=0.0cm 7.5cm 0.0cm 8.5cm, clip=true, width=0.75\columnwidth]{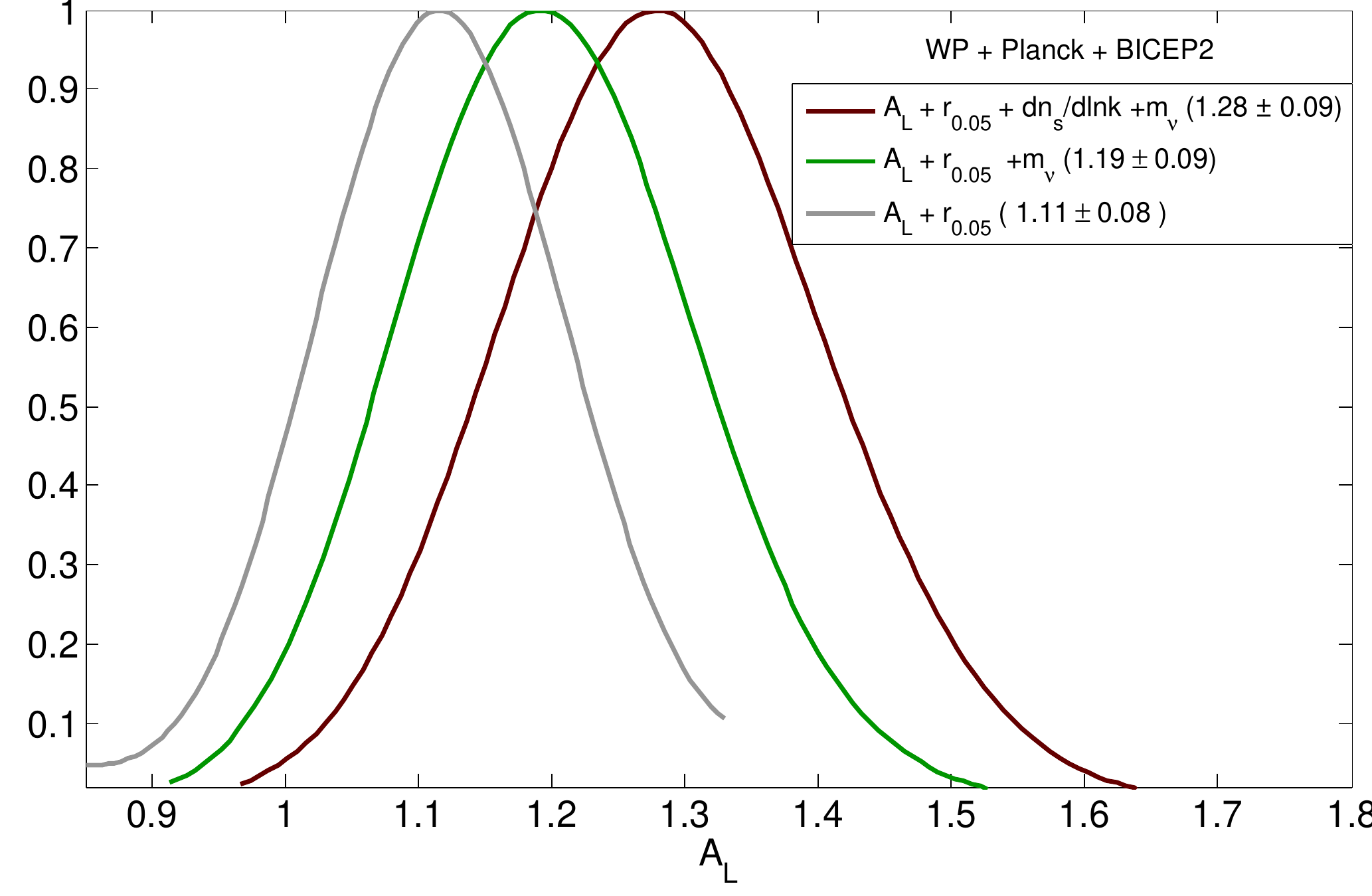}
\caption{\label{fig:Al}The brown curve is for a case where we vary $r_{0.05}$, $\sum m_{\nu}$,
$A_{L}$ and $dn_{s}/d\ln k$ along with the other standard
model parameters. We use WP+ Planck+ BICEP 2 likelihood, with prior on $\sum m_{\nu}$ from BOSS experiments. In the green curve we fix
the value of $dn_s/d\ln k=0$ and in the gray curve we fix
both $dn_s/d\ln k=0$ and $\sum m_{\nu}=0$. The mean value and the standard deviations
of $A_{L}$ are shown inside the bracket in the legend. }
\end{figure}

In this section we show the constraints on the $A_{L}$, i.e. the lensing power amplitude 
relative to the physical value \cite{Planck_param}. Theoretically, $A_{L}$ should be consistent with $1$.
However, Planck found that $A_{L}$ is inconsistent with its theoretical value at $2\sigma$ level. Here we consider the effect on $A_L$ due to the recent measurements of $\sum m_\nu$ 
and $r_{0.05}$. We vary $A_{L}$, $dn_{s}/d\ln k$, $\sum m_{\nu}$ and $r_{0.05}$ along with 
the other standard model parameters. The results are shown in Fig. \ref{fig:scatteredplot} and Fig. \ref{fig:Al}.
% We use the prior on $\sum m_\nu$  and $r_{0.05}$ as mentioned by  Beutler et. al. \cite{SDSS_1} and Ade et al. \cite{BICEP_1} respectively.

In Fig. \ref{fig:scatteredplot}
we show the scattered plot between $A_{L}$ and $dn_{s}/d\ln k$
and color coded with $\sum m_{\nu}$, which shows for $A_L\approx 1$ is consistent for the $r_{0.05}=0.2$ with lower value of $\sum m_\nu$ and without running.  It shows $A_{L}$ is negatively correlated with $dn_{s}/d\ln k$ and positively correlated with $r_{0.05}$. There is a mild positive correlation with correlation coefficient $0.06$ between $\sum m_\nu$ and $A_L$.  The one dimensional
marginal probability distribution for different cases are shown in
Fig. \ref{fig:Al}. It can be seen that if $dn_s/d\ln k=0$
and $\sum m_{\nu}=0$ then $A_{L}$ is consistent
with the physical value $A_L=1$ at $\sim2\sigma$ with the BICEP 2 likelihood\cite{BICEP_1}.  However, if we vary $dn_{s}/d\ln k$ and $\sum m_{\nu}$ then the average
value of $A_{L}$ shifts towards higher value. For the case with $SP+ r_{0.05}+ \sum m_\nu + dn_s/dk +A_L$ is 
inconsistent with $1$ at $3.1 \sigma$. This shows that varying $dn_{s}/d\ln k$ with the prior on 
$\sum m_\nu $ and $r_{0.05}$ leads to an inconsistent lensing amplitude, whereas model 
without $dn_{s}/d\ln k$ is inconsistent at $2.1 \sigma$. So, the model without running with the 
given $\sum m_\nu$ and $r_{0.05}$ is slightly preferred.  With the decrease in  $r_{0.05}$ from the value measured by BICEP 2, $A_{L}$ also decreases (as shown in Fig. \ref{fig:scatteredplot}). This will lead to a more consistent model with running spectral index and $\sum m_{\nu} = 0.36$.
\section{Discussions and Conclusions}\label{con}
%With several space and ground based missions for Cosmic Microwave Background (CMB), 
%we are now in an era of precision cosmology. Planck measurements matches well with the 
%$\Lambda$CDM model with the tensor to scalar ratio $r<0.11$. Similarly, Planck also put 
%a upper bound on mass of neutrino $\sum m_\nu <0.23$ eV. However, recent results from 
%BICEP 2 \cite{BICEP_1} reveals non-zero value of $r_{0.05}$ from $C_l^{BB}$ polarization spectra. 
%SDSS-III BOSS \cite{SDSS_1} also shows a signature of $\sum m_\nu= 0.36 \pm 0.10$ eV. 
%This development after the Planck results need to incorporate in estimating the cosmological 
%parameters.
%In this paper, 
We study the effects of the new measurements of  $r_{0.05}$ and $\sum m_\nu$ claimed by BICEP 2 and BOSS respectively, 
impose on other cosmological parameters. 
%Using the prior on $r_{0.05}$ and $\sum m_\nu$ as obtained from BICEP 2 \cite {BICEP_1} and BOSS \cite{SDSS_1} respectively, we estimated the best-fit Standard $6$ Parameters (SP), $\{\Omega_{b}h^{2},\Omega_{m}h^{2},h,\tau,n_{s},A_{s}\}$.
%using the cosmological parameter estimation code SCoPE \cite{scope}. 
%In Fig. \ref{fig:SP_nolensing} 
%and Fig. \ref{fig:SP_lensing}, we showed $4$ different cases, (i) SP (ii) SP + $\sum m_\nu$ (iii) 
%$SP + $r_{0.05}$ and (iv) SP + $\sum m_\nu$ +$r_{0.05}$. 
%Among all the $6$ parameters, $h$ shows  significant 
%variation. The model with $r_{0.05}$ and $\sum m_\nu =0$ increases the value of $h$, but with 
% $\sum m_\nu \neq 0$, the values of $h$ decreases significantly. 
%On estimating models with 
We evaluate the models  $dn_s/d\ln k$ and $N_{\rm eff}$ along with the other SP $\{\Omega_{b}h^{2},\Omega_{m}h^{2},h,\tau,n_{s},A_{s}\}$ for four different cases as mentioned in Sec. \ref{cp}. Our results show that  $dn_{s}/d\ln k \neq 0$ at $2.7\sigma$ (Sec. \ref{ns}) and $N_{\rm eff}$ is 
consistent with the theoretical value of $3.046$ within $1.74\sigma$ ( Sec. \ref{neff}). 
This implies that the simplest inflationary model and with the known relativistic species in the universe, explains the observed temperature anisotropy spectra of CMB. 

However, lensing amplitude $A_L$ which plays an important 
role in estimating the value of both $N_{\rm eff}$ and $dn_{s}/d\ln k$ does not gives a consistent estimation when considered running spectral index along with the current bound on $r_{0.05}$ and $\sum m_{\nu}$. 
%We have shown in Sec. \ref{al} that for 
The model with $\sum m_\nu$, $r_{0.05}$ and $dn_{s}/d\ln k =0$ makes $A_L$ is consistent 
with $1$, at $2.1 \sigma$. But %on considering the model with 
if we allow $dn_{s}/d\ln k$ to vary, then we get $A_L>1$ at 
$3.1\sigma$ level. %Any model with higher value of $A_L$ is highly unphysical and incorrect. 
Therefore, the models with running spectral index and the value of $r_{0.05}$ and $\sum m_{\nu}$ as measured by BICEP 2 and BOSS leads to an inconsistent cosmological model, %of measured $r_{0.05}$ and $\sum m_\nu$ 
%is not physical. %a correct  model.%From the above discussion 

Hence, we can conclude that the standard cosmological model with $SP+ \sum m_\nu +r$, 
$N_{\rm eff}=3.046$ and $A_L=1$, 
%consistent within $1\sigma$ and $2.67\sigma$ respectively and 
without running of the spectral index, $dn_{s}/d\ln k =0$ is fully consistent with the data. 
%Though it  from HST \cite{snia}. 
Though, it implies $H_0= 64.5 \pm 1.4$ kms$^{-1}$ Mpc$^{-1}$, that disagrees with supernova measurements ($H_0= 73.8 \pm 2.4$ kms$^{-1}$ Mpc$^{-1}$) from HST \cite{snia} by $6.6\sigma$. Improvement of measurements of $H_0$ and 
$\sum m_\nu$  from different future experiments can resolve this discrepancy. 
However, if Planck measures $r_{0.05}<0.2$, then the inconsistency arising due to lensing amplitude $A_L >1$ at more than $3 \sigma$ goes away (as shown in Fig. \ref{fig:scatteredplot}) and  the model with $\sum m_{\nu}$ and $dn_s/d\ln k$ becomes consistent. %significantly differs from the measurement of  % However, the other possible model is for lower value of  $\sum m_\nu$. For which, the agreement with $H_0$ improves, and it becomes consistent with supernova measurements at $2.3\sigma$. 
%$A_L=1$ and $N_{\rm eff}$ both remains consistent with the theoretical values. Running of spectral index is also consistent with zero in this model. 
In Table \ref{tab_3}, we summarize all these findings %results these two new measurements on cosmological parameters
 to give a comprehensive understanding of this analysis.
The constraints on $r_{0.05}$ can be improved with the full sky measurement of polarization data from Planck, which is expected to be released in near future. All these together can lead to more precise measurement of the cosmological models.
%Also, improvement of measurements of $H_0$ and $\sum m_\nu$  from different future experiments can lead to more precise measurement of the cosmological models.

%Our estimation of cosmological parameters will be improved further with this measurement. 
\begin{table}[h]
\centering
\caption{Summary for our analysis with WP+Planck+BICEP 2 likelihood for models beyond standard $6$ parameters ($\{\Omega_{b}h^{2}$, $\Omega_{m}h^{2}$, $h$, $\tau, n_{s}$, $A_{s}\}$) are mentioned below. }
\label{tab_3} 
\vspace{0.5cm}
\begin{tabular}{|p{9cm}|c|c|c|}
%\begin{tabular}{|l|l|||}
%\begin{tabular}{|clc|cl|p{8cm}|}
\hline 
%Model & Parameter  & BOSS  & BICEP 2 & BOSS + BICEP 2 \tabularnewline
Model & Parameter  & WP+ Planck+ BICEP 2 \tabularnewline
\hline
%SP & h & $0.640 \pm 0.014$ & $0.681 \pm 0.011$ & $0.644 \pm 0.013$ \tabularnewline
SP $+ \sum m_{\nu} + r_{0.05}$  & h &  $0.645 \pm 0.014$ \tabularnewline
\hline
%SP + $dn_{s}/d\ln k$ & $dn_{s}/d\ln k$ & -0.0083$\pm$0.0072 & -0.0198$\pm$ 0.0073 & -0.0147$\pm$0.0078\tabularnewline
SP $+ dn_{s}/d\ln k + \sum m_{\nu} + r_{0.05}$ & $dn_{s}/d\ln k$ & -0.0246$\pm$0.0091\tabularnewline
\hline
%SP + $N_{\rm eff}$ & $N_{\rm eff}$ & $3.0416_{\,}\pm0.2834$ & $3.5823 \pm 0.33$ & $3.2769 \pm 0.313$\tabularnewline
SP  $+N_{\rm eff} + \sum m_{\nu} + r_{0.05}$ & $N_{\rm eff}$ & $3.7530_{\,}\pm0.26$\tabularnewline
\hline
%SP + $A_{L}$ + $dn_{s}/d\ln k=0$& $A_L$ & -  & $1.15 \pm 0.08$ & $1.24 \pm 0.09$ \tabularnewline
SP + $A_{L}$ + $dn_{s}/d\ln k= 0$ + $\sum m_{\nu} + r_{0.05}$ & $A_L$ & $1.19 \pm 0.09$ \tabularnewline
\hline
%SP + $A_{L}$ + $dn_{s}/d\ln k \neq 0$ & $A_L$ & - & - & $1.36 \pm 0.11$ \tabularnewline
SP + $A_{L}$ + $dn_{s}/d\ln k \neq 0$ $+ \sum m_{\nu} + r_{0.05}$ & $A_L$ & $1.28 \pm 0.09$ \tabularnewline
\hline

%\hline 
\end{tabular}
\end{table}
%\pagebreak[4]
\acknowledgments{
We have used the HPC facility at IUCAA for the required computation.
S. D. and S. M. acknowledge Council for Science and Industrial Research
(CSIR), India, for the financial support as Senior Research Fellows.}


\begin{thebibliography}{99}
\bibitem{Planck_1}
 P. A. R.  Ade et al., Planck 2013 results. XV. CMB power spectra and likelihood, preprint arXiv:1303.5075v2, (2013).
 \bibitem{Planck_param}
P. A. R.  Ade et al., Planck 2013 results. XVI. Cosmological parameters, preprint arXiv:1303.5076v2, (2013).
\bibitem{BICEP_1}
P. A. R. Ade et al., BICEP 2 I: Detection Of B-mode Polarization at Degree Angular Scales, preprint arXiv:1403.3985, (2014).
\bibitem{SDSS_1}
F. Beutler et al. The clustering of galaxies in the SDSS-III Baryon Oscillation Spectroscopic Survey: Signs of neutrino mass in current cosmological datasets, preprint arXiv:1403.4599, (2014)

%F. Beutler et al. preprint arXiv:1403.4599, (2014).
\bibitem{Archidiacono2014}
M. Archidiacono, N. Fornengo, S. Gariazzo, C. Giunti, S. Hannestad, et al. Light sterile neutrinos after BICEP 2, arXiv:1404.1794v2, (2014).

\bibitem{Zhang2014}
J.-F. Zhang, Y.-H. Li and X. Zhang, Cosmological constraints on neutrinos
after BICEP 2, arXiv:1404.3598v1, (2014).

\bibitem{Zhang2014a}
J.-F. Zhang, Y.-H.  Li and X. Zhang, Sterile neutrinos help reconcile the observational
results of primordial gravitational waves from Planck and BICEP 2, arXiv:1403.7028v2, (2014).

\bibitem{Cheng2014}
C. Cheng and Q.-G. Huang, The Tilt of Primordial Gravitational Waves Spectra from
BICEP 2, arXiv:1403.5463v1, (2014).

\bibitem{Cheng2014a}
 C. Cheng and Q.-G. Huang, Constraints on the cosmological parameters from BICEP 2, Planck and WMAP, arXiv:1403.7173, (2014).

\bibitem{Cheng2014b}
C. Cheng, Q.-G. Huang and W. Zhao, Constraints on the extensions to the
base $\Lambda$CDM model from BICEP 2, Planck and WMAP, arXiv:1404.3467v1, (2014).

\bibitem{xia}
J. Q. Xia, Y. F. Cai, H. Li, X. Zhang, Evidence for bouncing evolution before inflation after BICEP 2, preprint arXiv:1403.7623, (2014).
\bibitem{Wu}
F. Wu, Y. Li, Y. Lu and X. Chen, Cosmological parameter fittings with the BICEP 2 data, preprint arXiv:1403.6462v1, (2014).

\bibitem{jerome}
J. Martin, C. Ringeval, R. Trotta, V. Vennin, Compatibility of Planck and BICEP 2 in the Light of Inflation, preprint arXiv:1405.7272, (2014).
\bibitem{scope}
S. Das and T. Souradeep, SCoPE: An efficient method of Cosmological Parameter Estimation, preprint arXiv:1403.1271, (2014).

\bibitem{camb}
A. Lewis and A. Challinor, CAMB code, \url{http://camb.info} .

%\bibitem{lewis}
%A. Lewis and S. Bridle, Cosmological parameters from CMB and other data: A Monte Carlo approach, Phys.Rev.D66, 103511, (2002).
 
\bibitem{wmap_like}
\url{"http://lambda.gsfc.nasa.gov/product/map/dr5/likelihood_get.cfm"}.

\bibitem{Planck_like}
\url{"http://wiki.cosmos.esa.int/planckpla/index.php/Main_Page"}.

\bibitem{bicep_like}
\url{http://bicepkeck.org/}

\bibitem{snia}
A.G. Riess, L. Macri, S. Casertano, H. Lampeitl, H.C. Ferguson, et al. A 3 \% Solution: Determination of the Hubble Constant with the Hubble Space Telescope and Wide Field Camera 3, Astrophys.J., 730, 119 (2011)

%\bibitem{con_1}
%R. L. Davis, H. M. Hodges, G. F. Smoot, P. J. Steinhardt, M. S. Turner, Cosmic microwave background probes models of inflation, Phys. Rev. Lett., 69, 1856,  (1992).
%\bibitem{con_2}
%A. R. Liddle and D. H. Lyth, COBE, gravitational waves, inflation and extended inflation, Phys. Lett., B., 291, 391,  (1992).
%\bibitem{con_3}
%J. R. Bond,  in Sasaki M., ed., Proc. 8th Nishinomiya-Yukawa Memorial Symposium, Relativistic Cosmology. Universal Academy Press, Tokyo, p. 23, (1994).
\bibitem{kosowsky}
A. Kosowsky and M. S. Turner, CBR anisotropy and the running of the scalar spectral index, Phys. Rev. D 52, R1739(R), (1995).

\bibitem{Gerbino2014}M. Gerbino, A.Marchini, L. Pagano, L. Salvati, E. Di Valentino,  and A. Melchiorri,  Blue Gravity Waves from BICEP 2 ?, arXiv:1403.5732v1, (2014).

\bibitem{Smith2014}K. M. Smith, C. Dvorkin, L. Boyle, N. Turok, M. Halpern,  G. Hinshaw and B. Gold, On quantifying and resolving the BICEP 2/Planck tension over gravitational waves, arXiv:1404.0373v1  (2014).

\end{thebibliography}
\end{document}